\documentclass[iop]{emulateapj}

\begin{document}
\title{A Three-Dimensional View of Turbulence: Constraints on Turbulent Motions in the HD 163296 Protoplanetary Disk using DCO$^+$}
\author{
Kevin M. Flaherty\altaffilmark{1},
A. Meredith Hughes\altaffilmark{1},
Sanaea C. Rose\altaffilmark{2},
Jacob B. Simon\altaffilmark{3,4},
Chunhua Qi\altaffilmark{5},
Sean M. Andrews\altaffilmark{5},
{\'A}gnes K{\'o}sp{\'a}l\altaffilmark{6,7}
David J. Wilner\altaffilmark{5},
Eugene Chiang\altaffilmark{8,9},
Philip J. Armitage\altaffilmark{4,10},
Xue-ning Bai\altaffilmark{5}}
\altaffiltext{1}{Van Vleck Observatory, Astronomy Department, Wesleyan University, 96 Foss Hill Drive, Middletown, CT 06459}
\altaffiltext{2}{Wellesley College, Wellesley, MA 02481}
\altaffiltext{3}{Department of Space Studies, Southwest Research Institute, Boulder, CO 80302}
\altaffiltext{4}{JILA, University of Colorado and NIST, 440 UCB, Boulder, CO 80309}
\altaffiltext{5}{Harvard-Smithsonian Center for Astrophysics, 60 Garden Street, Cambridge, MA 02138}
\altaffiltext{6}{Konkoly Observatory, Research Centre for Astronomy and Earth Sciences, Hungarian Academy of Sciences, Konkoly Thege Mikl{\'o}s {\'u}t 15-17, H-1121 Budapest, Hungary}
\altaffiltext{7}{Max-Planck-Institut f{\"u}r Astronomie, K{\"o}nigstuhl 17, D-69117, Heidelberg, Germany}
\altaffiltext{8}{Department of Earth and Planetary Science, 307 McCone Hall, University of California, Berkeley, CA 94720}
\altaffiltext{9}{Department of Astronomy, 501 Campbell Hall, University of California, Berkeley, CA 94720}
\altaffiltext{10}{Department of Astrophysical and Planetary Sciences, University of Colorado, Boulder, CO 80309-0391}
\begin{abstract}

Gas kinematics are an important part of the planet formation process. Turbulence influences planetesimal growth and migration from the scale of sub-micron dust grains through gas-giant planets. Radio observations of resolved molecular line emission can directly measure this non-thermal motion and, taking advantage of the layered chemical structure of disks, different molecular lines can be combined to map the turbulence throughout the vertical extent of a protoplanetary disk. Here we present ALMA observations of three molecules (DCO$^+$(3-2), C$^{18}$O(2-1) and CO(2-1)) from the disk around HD 163296. We are able to place stringent upper limits ($v_{\rm turb}<$0.06c$_s$, $<$0.05c$_s$ and $<$0.04c$_s$ for CO(2-1), C$^{18}$O(2-1) and DCO$^+$(3-2) respectively), corresponding to $\alpha\lesssim$3$\times$10$^{-3}$, similar to our prior limit derived from CO(3-2). This indicates that there is little turbulence throughout the vertical extent of the disk, contrary to theoretical predictions based on the magneto-rotational instability and gravito-turbulence. In modeling the DCO$^+$ emission we also find that it is confined to three concentric rings at 65.7$\pm$0.9 au, 149.9$^{+0.5}_{-0.7}$ au and 259$\pm$1 au, indicative of a complex chemical environment. 

\end{abstract}

\section{Introduction}
The planet formation environment in disks around young stars is highly dynamic, subject to large azimuthal and radial velocities, coupled gas and dust dynamics, as well as perturbations from magnetic fields, embedded planets and/or passing stars. While these phenomena are challenging to fully characterize, the Atacama Large Millimeter/submillimeter Array (ALMA), with its ability to spatially resolve doppler motions from multiple molecular emission lines at high signal-to-noise, is advancing our understanding of the protoplanetary disk kinematic structure. ALMA observations have used the dominant Keplerian motion to measure the mass of the central star(s) \citep{cze15,cze16}, while deviations from the Keplerian rotation have been investigated for warps \citep{cas15}, fast radial streams \citep{ros14,van17}, winds \citep{sal14}, gas pressure support, self-gravity \citep{ros13} and even proto-planetary candidates \citep{fac17}. 

The amplitude and nature of the motion within the disk represent important parameters in the planet formation process. In particular, non-thermal motions can influence processes ranging from the collisional growth of small dust grains \citep{tes14} to the ability of massive planets to open a gap \citep[e.g.][]{fun14}. Because of these effects, constraints on the strength of turbulence, especially near the disk midplane where planet formation is likely to occur, are crucial for understanding the evolution of planets and planetary systems.

Constraints on the vertical structure of turbulence are also valuable for understanding the physical mechanism driving turbulence. The magneto-rotational instability \citep[MRI][]{bal91,bal98}, which in protoplanetary disks relies on the coupling of magnetic fields to weakly ionized gas in a rotating disk, generates turbulence that can increase from a few percent of the local sound speed near the midplane up to nearly the sound speed in the upper layers \citep{fro06,sim13b,sim15}. Gravito-turbulence, which relies on gravitational instabilities near the midplane, also predicts large non-thermal motions, but with a nearly constant vertical profile \citep{for12,shi14}. The Vertical Shear Instability may also produce turbulence at the level of a few percent of the sound speed \citep{nel13}, although the vertical structure of the turbulence arising from this instability has not been fully characterized in detail.  

Previous radio observations using the SMA and PdBI have found tentative evidence for turbulence \citep[e.g.][]{hug11,gui12}. In our analysis of ALMA Science Verification data of the disk around HD 163296 \citep{fla15}, we placed a tight constraint on the turbulence in the upper layers ($v_{\rm turb}<$0.04c$_s$) using CO(3-2) emission, but the modest signal-to-noise of the more optically thin $^{13}$CO(2-1) and C$^{18}$O(2-1) emission prevented us from tightly constrainting the motion near the midplane. \citet{tea16} use CO(2-1), CN(2-1) and CS(2-1) to measure turbulence in the disk around TW Hya, finding non-thermal motion from the CO(2-1) emission layer, although the uncertainty in the amplitude calibration severely limit the measurements.

We take advantage of the complex chemical environment within protoplanetary disks to probe the vertical structure of turbulence. CO is the second most abundant molecule after H$_2$ and is highly optically thick. The emission from less abundant isotoplogues, such as $^{13}$CO or C$^{18}$O, originates deeper into the disk, but the ability of these isotopologues to trace the midplane is still limited by the freeze-out of CO gas onto dust grains \citep[e.g.][]{qi15}. Other molecules may remain abundant in the gas phase at colder temperatures. DCO$^+$ emission, which has been observed in a number of protoplanetary disk systems \citep{van03,gui06,qi08,obe10,obe11,mat13,tea15,qi15,yen16,hua17}, is predicted to arise from close to the cold midplane \citep{wil07,obe12,obe15,tea15}. Recent observations have found that DCO$^+$ may exist in regions of the disk where CO was previously believed to be frozen out \citep{obe15}. This spatial distribution makes it a particularly useful tool for measuring gas kinematics in a way that complements the much brighter CO emission. 

Here we present an analysis of ALMA observations of DCO$^+$(3-2), C$^{18}$O(2-1) and CO(2-1) from the disk around the Herbig star HD 163296. \citet{fla15} previously found $v_{\rm turb}<$0.04c$_s$ in the upper layers of this system based on the ALMA Science Verification CO(3-2). With these new high S/N observations we can probe similar levels of non-thermal motion for a region closer to the midplane, where planet formation is expected to occur. In section 2 we present our data, in section 3 we lay out an initial analysis of the structure, which informs our detailed model (section 4) and the subsequent results (section 5). In section 6 we discuss the implications of the weak turbulence seen in all three tracers.

\section{Data\label{data}}
Cycle 2 observations (2013.1.00366.S) with the Atacama Large Millimeter/submillimeter Array of HD 163296 were taken over the course of five nights (2014 June 4, 14, 16, 17 and 29). During each night the on-source integration time was 52 minutes, and the nearby quasar J1733-1304 was used for phase and gain calibration. Baselines ranged from 10k$\lambda$ to 480k$\lambda$, corresponding to a spatial resolution of $\sim0\farcs5$. 

Four spectral windows were observed: spectral window 0 was centered on CO(2-1), spectral window 2 was centered on DCO$^+$(3-2), spectral window 3 was centered on C$^{18}$O(2-1) and spectral window 1 was a continuum window centered at 232.71 GHz. The spectral windows centered on line emission had 3840 channels, each 0.015 MHz (20 m s$^{-1}$) wide. The continuum spectral window had a total bandwidth of 2 GHz covered by 128 channels. Data were self-calibrated using the continuum spectral window. 

The spectral line data were continuum-subtracted and binned by a factor of 10, to 0.2 km s$^{-1}$ wide channels, to increase S/N and decrease the time taken to model the emission. We do not anticipate that this spectral binning will substantially limit our ability to constrain turbulence; we found similar constraints on turbulence based on CO(3-2) emission when using either a 0.1 km s$^{-1}$ resolution spectrum ($<$0.04c$_s$) or a 0.3 km s$^{-1}$ spectrum ($<$0.06c$_s$) \citep{fla15}. Spatial resolution and S/N play a larger role in our ability to tightly constrain turbulence in the outer disk. Uncertainties on the visibilities were derived based on the dispersion around each baseline in line-free channels. Images were generated using robust weighting for the spectral lines and uniform weighting for the continuum window, resulting in beam sizes of $\sim0\farcs5$ and $\sim0\farcs4$ respectively. A clean mask based on the emission pattern of a Keplerian disk was applied to the CO(2-1) data. Additionally, we extracted DCO$^+$(5-4) and Band 7 continuum data from the ALMA Science Verification observations of HD 163296, with self-calibration applied based on the HD 163296 Band 7 Science Verification CASA guide\footnote{https://almascience.nrao.edu/alma-data/science-verification}.

Derived quantities for the cycle 2 observations, which reach peak S/N of 50-1000, are listed in Table~\ref{data_table}. \citet{ros13} measure CO(2-1) and C$^{18}$O(2-1) fluxes of 46$\pm$5 Jy km s$^{-1}$ and 5.8$\pm$0.6 Jy km s$^{-1}$ respectively from the ALMA science verification data. \citet{qi11} report fluxes for these same lines of 54.17$\pm$0.39 Jy km s$^{-1}$ and 6.40$\pm$0.16 Jy km s$^{-1}$ based on SMA observations. \citet{qi15} and \citet{bon16} report C$^{18}$O(2-1) integrated fluxes from the ALMA Science Verification data of 6.8$\pm$0.7 Jy km s$^{-1}$ and 6.2$\pm$0.4 J km s$^{-1}$. SMA observations of the 1.3mm continuum by \citet{ise07} measure a flux of 705$\pm$12 mJy. ALMA observations of DCO$^+$(3-2) have been reported by \citet{yen16} and \citet{hua17}, with \citet{hua17} deriving an integrated line flux of 1.29$\pm$0.04 Jy km s$^{-1}$. Our integrated line and continuum intensities are consistent with these previous measurements, when accounting for uncertainties in calibration, which are much larger than the statistical uncertainties. The flux of the amplitude calibrator source varied from 1.211 $\pm$ 0.001 Jy to 1.313 $\pm$ 0.005 Jy over the course of our observations, while the continuum and individual lines vary by 5-10\%\ from night to night, suggesting larger errors due to night-to-night calibration. An additional uncertainty arises from the imprecise knowledge of the flux of the calibrator source; the ALMA calibrator database lists a 15\%\ uncertainty on the flux at 233 GHz of J1733. Throughout our analysis we assume a 20\%\ uncertainty on the amplitude calibration of our data.

\section{Initial DCO$^+$ Morphology Estimate\label{initial}}
\begin{figure*}
\includegraphics[scale=.4]{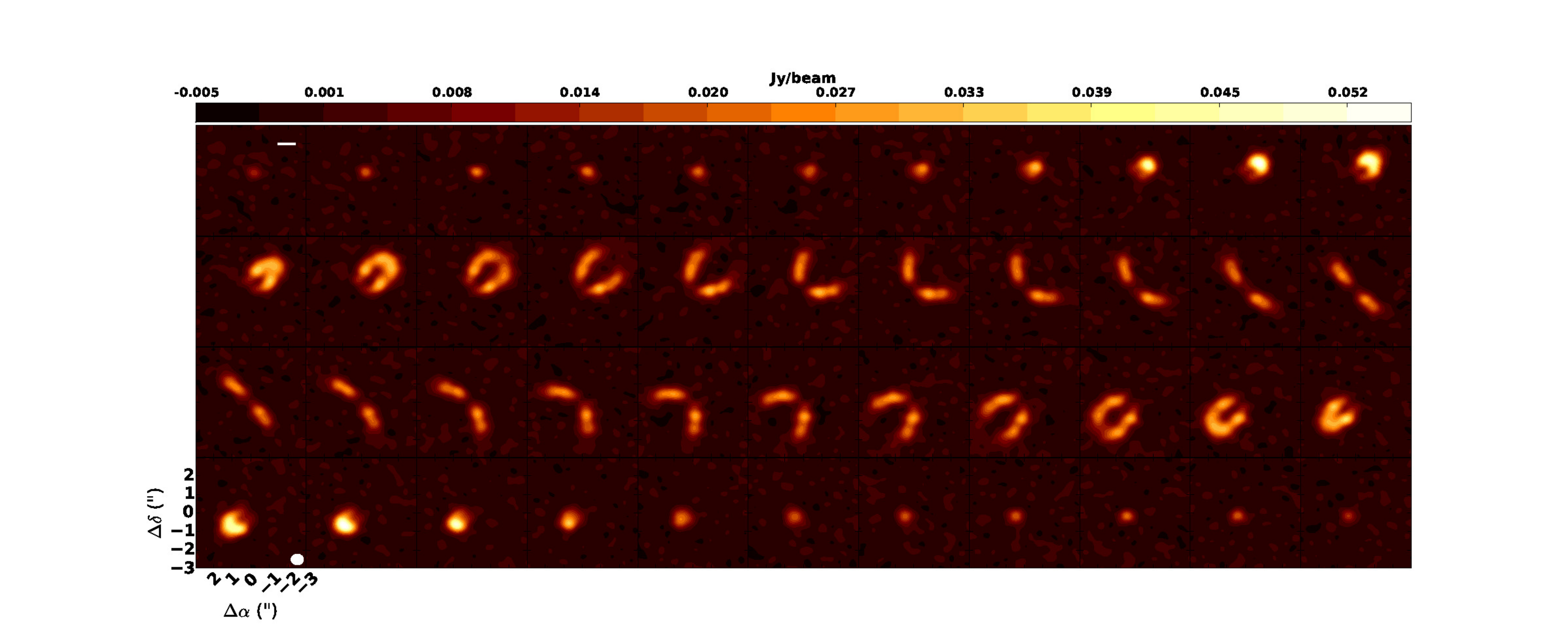}
\caption{Channel maps for DCO$^+$(3-2) emission. The beam shape is indicated by the circle in the lower left panel, while the white horizontal line in the upper left panels indicates 100 au. Emission is detected at high S/N throughout the emission line. The channels probing the outer disk exhibit a 'peanut' like structure that is characteristic of emission from two distinct rings, rather than a smooth density distribution.\label{chmaps_dcoplus}}
\end{figure*}

\begin{figure}
\includegraphics[scale=.4]{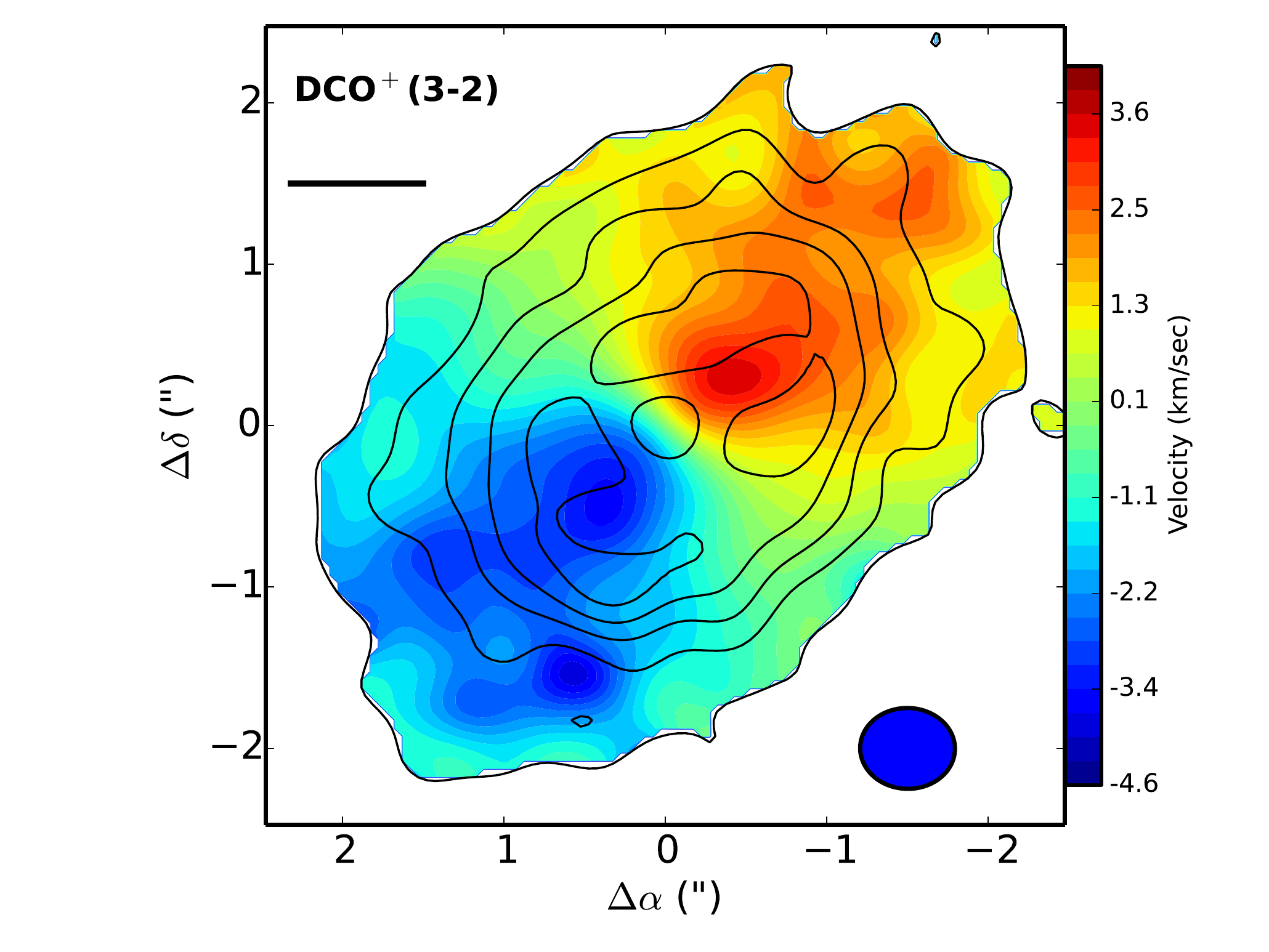}
\caption{Moments 0 (integrated intensity, contours) and 1 (intensity-weighted velocity, colors) for DCO$^+$(3-2). Contours are in steps of 3$\sigma$ ($\sigma$=3mJy km s$^{-1}$). Beam shape is indicated in the lower right, while the scale bar in the upper left indicates 100 au.\label{moments_dcoplus}}
\end{figure}

While previous observations of DCO$^+$ in the disk around HD 163296 found a broad ring \citep{mat13,qi15}, the high S/N and spatial resolution of our data are able to reveal new details about this structure (Figure~\ref{chmaps_dcoplus}). Emission is clearly detected out to $\sim$250 au (Figure~\ref{moments_dcoplus}) with peak S/N$\sim$50. The channels near the line peak reveal additional structure. Rather than smooth emission we see a peanut-shaped feature, which is suggestive of the presence of two distinct rings of DCO$^+$. Recent observations have begun to reveal gas rings around other stars in both DCO$^+$ \citep{obe15} and CO \citep{sch16,hua16} and such behavior here is not surprising, especially in light of the multi-ringed structure recently detected in the dust continuum emission at millimeter wavelengths \citep{zha16,ise16}.

To obtain an initial estimate of the structure, we employ an image deconvolution routine modeled after \citet{luc74} and \citet{sco83}. This procedure utilizes concentric rings of emission moving in Keplerian orbits to match the full three-dimensional (position-position-velocity) data set. By iteratively varying the emissivity of each ring independently until it matches the emission profile, while accounting for the finite spectral and spatial resolution, this process can fit an arbitrary radial distribution of gas. The known velocity profile, Keplerian to 1st order, allows for a deconvolution of structure even below the nominal spatial resolution ($\sim0\farcs5$=60 au). 

\begin{figure}
\includegraphics[scale=.3]{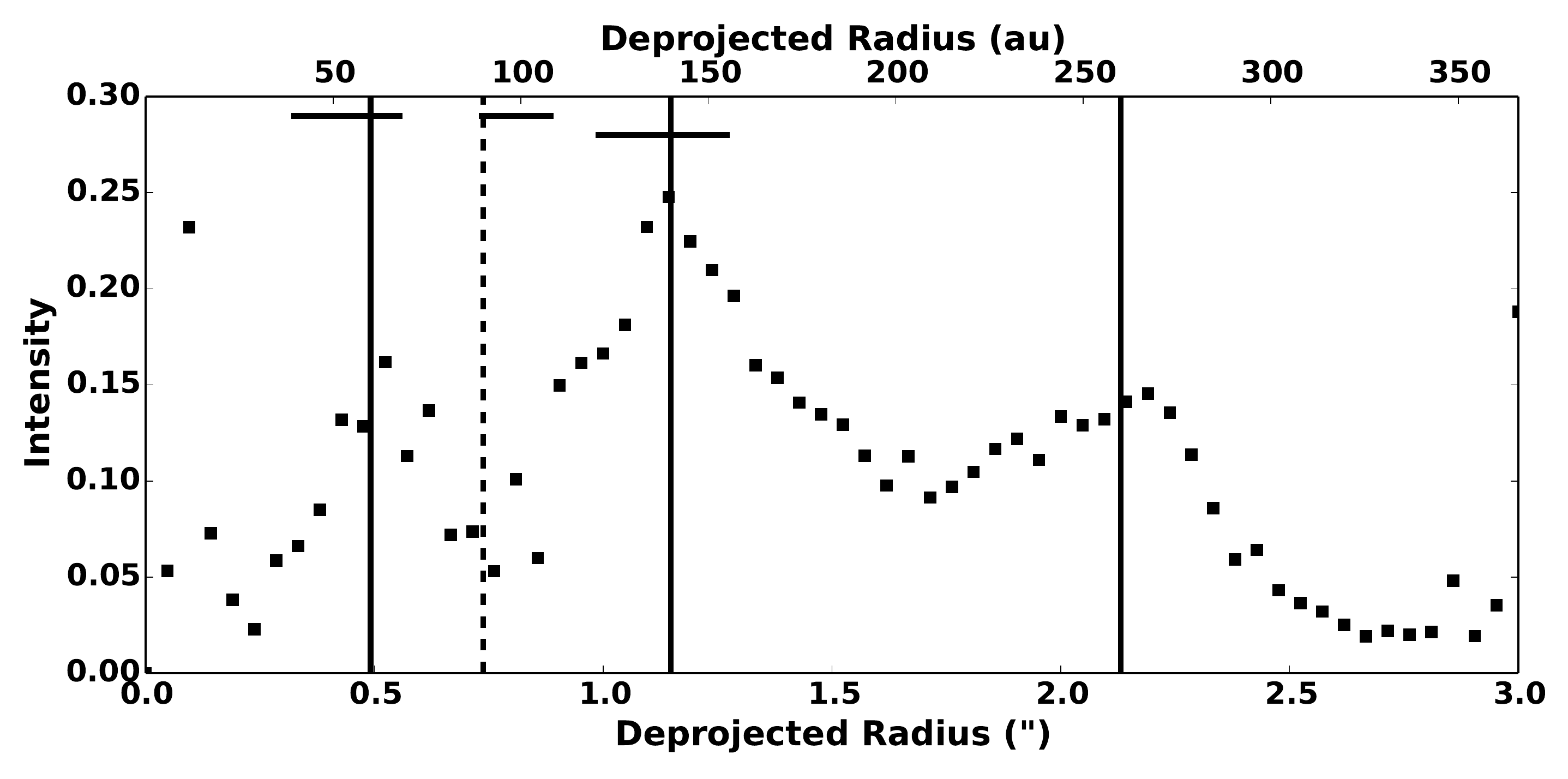}
\caption{Radial distribution of DCO$^+$(3-2) intensity (in arbitrary units) as derived by the Richardson-Lucy deconvolution algorithm. DCO$^+$(3-2) is composed of three rings, marked by vertical lines, at $\sim$60, $\sim$140 and $\sim$260 au. None of these rings are located at the position of the CO snow line (90 au, vertical dashed line), although there is some overlap between two of the DCO$^+$ rings and two of the gaps in the continuum emission (\citet{ise16}, horizontal solid lines).\label{deconvolution_dcoplus}}
\end{figure}

When applied to the DCO$^+$(3-2) data, this routine finds three distinct rings of emission at $\sim$60, $\sim$140 and $\sim$260 au (Figure~\ref{deconvolution_dcoplus}). The outer two rings were evident based on the appearance of the channel maps, while the third ring appears as enhanced emission within the line wings. The overall distribution of DCO$^+$, with emission covering $\sim$60 au to $\sim$250 au, is consistent with that found from previous studies. \citet{qi15} detect DCO$^+$(4-3) emission from 40 au to 250 au, while \citet{mat13} find DCO$^+$(5-4) in a broad ring centered at roughly 90 au. Our high S/N data is able to resolve the broad features seen in these previous data into its underlying constituents.

Deconvolution reveals details about the radial surface brightness structure, but does not constrain the density or temperature within the rings. The previous measurements of DCO$^+$(4-3) \citep{qi15} and DCO$^+$(5-4) \citep{mat13} emission can be combined with our DCO$^+$(3-2) measurement to obtain an initial estimate of the excitation temperature of the DCO$^+$ gas. The excitation temperature defined by the J=(u$_1$-l$_1$) and J=(u$_2$-l$_2$) transitions (u$_1$$>$u$_2$) is: 
\begin{equation}
T_{ex}=\frac{h \nu_{u_1u_2}}{k\ln(\frac{S_{u_1l_1}g_{u_2}A_{u_2l_2}\nu_{u_2l_2}}{S_{u_2l_2}g_{u_1}A_{u_1l_1}\nu_{u_1l_1}})},
\end{equation}
where $\nu_{u_1u_2}$ is the frequency of the transition between levels $u_1$ and $u_2$, $S_{u_1l_1}$ is the integrated flux density between the $u_1$ and $l_1$ levels, $g$ is the statistical weight and $A$ is the Einstein A value. Including a 20\% uncertainty on the amplitude calibration for each data set, we find a temperature of 16$\pm$4 K and 11$\pm$4 K when comparing DCO$^+$(3-2) to DCO$^+$(5-4) and DCO$^+$(4-3) respectively. These low temperatures suggest that the DCO$^+$ arises from the cold temperature midplane rather than the warm surface layers. This calculation assumes LTE, which is likely applicable for DCO$^+$. The DCO$^+$(3-2) and DCO$^+$(5-4) lines have critical densities of 1.9$\times$10$^{6}$ cm$^{-3}$ and 1.4$\times$10$^{7}$ cm$^{-3}$ respectively. According to the density and temperature structure derived in \citet{fla15}, typical densities of $\sim$10$^{9}$ cm$^{-3}$ are found near the midplane. Densities below 10$^{6}$ cm$^{-3}$ are only seen in the upper reaches of the outer disk ($z>$ 80 au for $r>$ 200 au), an area that does not contribute significantly to the total emission. The excitation temperature derived from the line intensity ratios is an intensity-weighted disk average temperature. In a realistic disk the temperature decreases with distance from the star and based on our previous analysis \citep{fla15} we expect the temperature to vary by $\sim$30\% between the inner and outer rings, which is comparable to the uncertainty in our temperature estimate.

\section{Disk Structure Model\label{model}}
In order to constrain turbulence from the line emission, we employ a parametric disk structure and ray tracing code to generate model visibilities that can be compared directly to the data. The modeling code comes from \citet{fla15}, which is based on earlier work by \citet{ros13} and \citet{dar03}. We provide a brief summary here with more detail available in \citet{fla15}.

The underlying surface density structure is assumed to follow the functional form predicted for a viscously evolving disk \citep{lyn74,har98}.
\begin{equation}
\Sigma_{\rm gas}(r) = \frac{M_{\rm gas}(2-\gamma)}{2\pi R^2_c}\left(\frac{r}{R_c}\right)^{-\gamma}\exp\left[-\left(\frac{r}{R_c}\right)^{2-\gamma}\right].
\end{equation}
Here $M_{\rm gas}$ is the total gas mass, $R_c$ is the critical radius and $\gamma$ controls the power law shape of surface density for $r\ll R_c$. The temperature is assumed to follow a power law with radius, with a vertical gradient connecting the cold midplane ($T_{\rm mid}$) with the warm atmosphere ($T_{\rm atm}$). 
\begin{eqnarray}
T_{\rm mid} = T_{\rm mid0}\left(\frac{r}{150\ \rm au}\right)^{q}\\
T_{\rm atm} = T_{\rm atm0}\left(\frac{r}{150\ \rm au}\right)^{q}\\
T_{\rm gas}(r,z) = \left\{
\begin{array}{ll}
T_{\rm atm} + (T_{\rm mid}-T_{\rm atm})(\cos\frac{\pi z}{2Z_q})^{2} & \mbox{if $z < Z_q$} \\
T_{\rm atm} & \mbox{if $z \ge Z_q$}
\end{array}
\right.\\
Z_q = 70\ {\rm au} (r/150\ {\rm au})^{1.3}
\end{eqnarray}
The hydrostatic equilibrium calculation is performed to derive the volume density given the temperature and surface density structure. The velocity field is Keplerian, with corrections for the height above the midplane and the pressure support of the gas, as in \citet{ros13}. The line is assumed to be a Gaussian with width
\begin{equation}
\Delta V = \sqrt{\left(2k_BT(r,z)/m_{CO}\right) + v_{\rm turb}^2}.
\end{equation}
We included a non-thermal line broadening term, $v_{\rm turb}$, which is assumed to be proportional to the local isothermal sound speed ($\sqrt{2k_BT(r,z)/\mu m_{H}}$). We interpret this additional broadening as due to turbulence, although other deviations from Keplerian and thermal motion may contribute to this term. 

Our main goal is to use the individual lines to constrain the turbulent structure at the various vertical emission layers of these molecules within the disk. As such, we start with similar underlying structures, but allow for variations between the molecules to best fit the data. While there is certainly a single temperature/density/kinematic structure that applies to all of the emission from this system, the assumptions needed to fit an individual line (e.g. assuming a temperature structure to break its degeneracy with surface density for an optically thin line), and our incomplete knowledge of the complex physics/chemistry in protoplanetary disks (e.g. selective photo-dissociation, CO non-thermal desorption), can lead to differences in model parameters derived from different lines. The constraints on the model parameters simply reflect the values needed, in the context of our assumed parametric model, to reproduce the flux from the emitting region of a particular line. 

With this framework in mind, we consider the detailed structure for each molecule in turn. CO is confined to a molecular layer, whose upper boundary is set by photodissociation and whose lower boundary is set by freeze-out. The photodissociation boundary is defined by the height at which the vertical column density reaches $\Sigma_{21}$=0.79, where $\Sigma_{21}$ is the surface density in units of 1.59$\times$10$^{21}$ cm$^{-2}$. Freeze-out occurs below a temperature of 19 K, and for CO, which is less sensitive to midplane abundance, we assume that freeze-out leads to a drop in CO abundance by eight orders of magnitude. For C$^{18}$O(2-1), which is more sensitive to the abundance at the cold midplane, we assume an abundance decrement of a factor of five, consistent with recent studies \citep{qi15}. We assume isotope abundances of $^{18}$O/$^{16}$O=557 and $^{13}$C/$^{12}$C=69 \citep{wil99}. The abundance of CO relative to H$_2$, $X_{\rm CO}$, is allowed to vary when modeling the optically thin C$^{18}$O, and is fixed at 10$^{-4}$ when modeling optically thick CO(2-1). We assume LTE for all models which, as discussed earlier, is a reasonable assumption even for DCO$^+$.


Since DCO$^+$ is optically thin, it is difficult to disentangle the density and temperature, and we choose to fix the temperature structure while allowing the density to vary. We use the temperature parameters derived in \citet{fla15} from the CO(3-2) emission ($q$=-0.216, $R_c$=194 au, $T_{\rm atm0}$=94 K, $\gamma=1$, $T_{\rm mid0}$=17.5 K). The uncertainty on these parameters will feed into the uncertainties on the density and turbulence, an effect that we consider in more detail below. The DCO$^+$ molecules are assumed to be distributed in three rings whose radial abundance profile is given by:
\begin{eqnarray}
\left[\frac{DCO^+}{H_2}\right] = \left[\frac{DCO^+}{H_2}\right]_{\rm in}\exp\left(-\frac{(r-R_{\rm in})^2}{\sigma_{\rm in}^2}\right)\nonumber\\
+\left[\frac{DCO^+}{H_2}\right]_{\rm mid}\exp\left(-\frac{(r-R_{\rm mid})^2}{\sigma_{\rm mid}^2}\right)\nonumber\\
+\left[\frac{DCO^+}{H_2}\right]_{\rm out}\exp\left(-\frac{(r-R_{\rm out})^2}{\sigma_{\rm out}^2}\right),
\end{eqnarray}
 where $R_{\rm in}$, $R_{\rm mid}$ and $R_{\rm out}$ are the central locations of the three rings and [DCO$^+$/H$_2$]$_{\rm in}$, [DCO$^+$/H$_2$]$_{\rm mid}$ and [DCO$^+$/H$_2$]$_{\rm out}$ are the peak abundances of DCO$^+$ in the three rings. We assume $\sigma_{\rm in}$, $\sigma_{\rm mid}$ and $\sigma_{\rm out}$ are equal to one-tenth of their respective central radii; we examine the consequences of this assumption later. The abundance is assumed to be constant between the vertical column densities $\Sigma_{21}=0.79,1000$. Due to the low excitation temperature derived above, the complex radial structure with emission at radii where CO (a key molecule in the creation of DCO$^+$) is expected be frozen out, and the evidence for modest depletion of CO gas during freeze out \citep{qi15} we exclude the effects of freeze-out from our DCO$^+$ model. Including DCO$^+$ freeze-out for T$<$19 K in an additional model fit, we find that only [DCO$^+$/H$_2$]$_{\rm mid}$ and [DCO$^+$/H$_2$]$_{\rm out}$ move from their fiducial values, increasing by 0.1 and 0.3 dex respectively. For our fiducial DCO$^+$ model the free parameters are $R_{\rm in}$, $R_{\rm mid}$, $R_{\rm out}$, [DCO$^+$/H$_2$]$_{\rm in}$, [DCO$^+$/H$_2$]$_{\rm mid}$, [DCO$^+$/H$_2$]$_{\rm out}$ as well as v$_{\rm turb}$ and the inclination.

Our final CO(2-1) models have free parameters $q$, $R_c$, $T_{\rm atm0}$, $T_{\rm mid0}$, $R_{\rm in}$, $v_{\rm turb}$, and inclination, while for C$^{18}$O(2-1) we fix $R_c$, $T_{\rm atm0}$, and $T_{\rm mid0}$ and add $X_{\rm CO}$ and $\gamma$ as free parameters. Allowing the temperature to vary when fitting CO(2-1), while fixing it for C$^{18}$O(2-1) and DCO$^+$ opens up the possibility of different temperature structures for the different molecules. After fitting, we confirm that the differences in temperature are small, especially compared to the underlying systematic uncertainty in amplitude calibration. After finding approximately accurate parameters we refine the center of the disk relative to the phase center as well as the velocity center of the line using a simple grid search. We find these to be [-0$\farcs$03,0$\farcs$02] and 5.74 km s$^{-1}$ consistently between all of the lines. The position angle is fixed (PA=312$^{\circ}$) but inclination is allowed to vary between the three molecules since its measurement will depend in part on the vertical location of the emission within the disk. As with the temperature, we find consistent results for all three molecules. We also assume a distance ($d$=122 pc), stellar mass \citep[$M_*$=2.3$M_{\odot}$;][]{mon09}, and disk mass ($M_{\rm gas}$=0.09$M_{\odot}$). While not included in our modeling, the $\sim$10\%\ uncertainty on the distance to HD 163296 \citep{van07} will directly translate into a $\sim$10\%\ systematic uncertainty on any radial distances measured in this system. The disk mass was taken from \citet{ise07}, who used continuum observations of sub-mm dust emission to derive a dust mass, and converted to a gas mass assuming a gas to dust mass ratio of 100. Recent work has brought into question standard assumptions about gas to dust mass ratio, CO/H$_2$, and C$^{18}$O/CO \citep{ber13,mio14,ans16,kam16}, raising the possibility of a significant deviation from our assumed disk mass. To accommodate this we allow $X_{\rm CO}$ to vary, but note that any deviations from the fiducial value of 10$^{-4}$ may reflect the fact that the assumed dust to gas ratio or $^{18}$O/$^{16}$O abundance is incorrect.

Once a suite of parameters has been specified, model images are generated and converted to visibilities using the MIRIAD task UVMODEL. The model visibilities are then directly compared to the data, with the goodness of fit calculated using the chi-squared statistic. To derive the posterior distributions for each parameter, we employ the MCMC routine EMCEE \citep{for13} based on the affine-invariant algorithm originally proposed in \citet{goo10}. Our chains typically consist of 80 walkers and 1000 steps, with the first 500 removed as burn-in. Simple, physical, priors are employed (e.g. $v_{\rm turb}>0$) and we employ linear spacing in all parameters except $R_c$, where we fit $\log(R_c)$. 

Systematic uncertainty in the amplitude calibration of the data is an important source of error since it dominates the statistical noise in these high S/N data. Since systematic uncertainty can directly affect the temperature derived from optically thick lines, and hence the thermal broadening, it potentially has a large effect on a derived non-thermal broadening \citep{tea16}, although high spatial resolution imaging can help to break this degeneracy \citep{sim15}. To account for systematic uncertainty in our MCMC modeling framework, we introduce an additional parameter, $sys$, that is a multiplicative scale factor on the amplitude of the model emission. The model is scaled down by $sys$ to force the other parameters (e.g. temperature, abundance, turbulence) to adjust upwards accordingly. A value of $sys=0.8$ corresponds to a true disk flux 20\%\ lower than is derived from our data, while $sys=1.2$ simulates a true disk flux 20\%\ higher than we derived from our data. By varying $sys$ between 0.8 and 1.2 we can sample the full size of the posterior distribution functions allowed by the systematic uncertainty.  

\subsection{The Influence of Model Assumptions}
Throughout our analysis we assume one functional form for the temperature and density structure, which is an approximation of more detailed models, which are themselves approximations of reality. The vertical temperature profile in particular is highly uncertain. While we employ the Type II profile from \citet{dar03} as the form of the vertical temperature structure, which is similar to the temperature structure derived in the radiative transfer models of \citet{dal06}, we could have similarly chosen the Type I profile, an exponential, which more closely matches the Herbig Ae disk radiative transfer models of \citet{jon07}. Analysis of prior observations have not strongly favored one model over the other, providing little guidance as to which model prescription to choose.

In the end these choices do not strongly influence our results, because each emission line is tracing a relatively narrow vertical region of the disk, and many different model structures can be made to pass through the constraints on temperature and density in this region. DCO$^+$(3-2) and C$^{18}$O(2-1) lie roughly within one pressure scale height of the midplane, while CO(2-1) and CO(3-2) emit from between 2 and 3 pressure scale heights above the midplane. With CO(2-1) and CO(3-2) we are able to spatially resolve the near and far side of the disk, which provides an additional constraint on the midplane temperature, but does not fully constrain the functional form of the temperature profile connecting the midplane and the surface layers. An equally acceptable fit can be found with a lower/higher $Z_{q0}$ than has been assumed here, as well as with an exponential temperature profile, with suitable adjustments to $T_{\rm atm0}$. Similarly, different choices for $\gamma$ can be accommodated with variations in $R_c$ when fitting optically thick lines like CO(2-1) and CO(3-2). The exact value of the individual parameters (e.g. $T_{\rm atm0}$, $R_c$) will vary with the model prescription, but the temperature and location of the emitting region of each molecule is consistently constrained across the different model prescriptions. In Section~\ref{dco+_models} we examine variations on the DCO$^+$ model structure and find that they do not substantially bias our turbulence measurement. A simultaneous fitting of multiple isotopologues and transitions is needed to more accurately constrain the shape of the temperature profile, and is beyond the scope of this paper. 

For turbulence we make the simplifying assumption that the constant of proportionality between the non-thermal motion and the local sound speed does not vary throughout the disk. If MRI is operating, then this is certainly not the case, as turbulence is expected to increase towards the surface layers of the disk. Here again the fact that each line probes are relatively narrow vertical region makes our assumption less severe. Across a single pressure scale height, variations in the non-thermal motion are typically less than a factor of two \citep{sim15}. Assuming ambipolar diffusion is the main non-ideal MHD effect, variations in the non-thermal motion are also less than a factor of two from 30 to 100 au \citep{sim15}. The lack of strong non-thermal motion in the CO(3-2) observations of the disk around HD 163296 \citep{fla15} also indicate that we are not yet justified in introducing additional parameters associated with a vertical or radial profile of turbulence, nor is there enough information to determine the correct parameterization of turbulence (ie. as a fraction of the local sound speed, or as a fixed velocity in units of km s$^{-1}$). In the end we are constraining a disk-averaged, intensity-weighted value of the non-thermal non-Keplerian motion, as a fraction of the local sound speed, for each molecule.

By treating turbulence as a line broadening term we have also ignored any influence non-thermal motion will have on the temperature structure. Strong turbulence will lift small dust grains high into the atmosphere \citep{dul04,fro06b,fro09}, changing the absorption of the stellar radiation, leading to higher temperatures at larger heights in the disk \citep{dal06}. In the context of our temperature functional form, significant dust settling due to weak turbulence would be similar to a small $Z_{q0}$ \citep{qi11}. As discussed above, an equally acceptable model can be found with small $Z_{q0}$ as with our assumed value of $Z_{q0}$, with a corresponding adjustment of $T_{\rm atm0}$. Turbulence can also influence the chemical structure; weak turbulence may lead to higher depletion of CO \citep{fur14,xu17} while strong turbulence may blur the sharp boundaries associated with CO freeze-out \citep{sem11,fur14}. While we do allow [CO/H$_2$] to vary in our models of C$^{18}$O(2-1), we do not adjust the sharpness of the CO freeze-out boundary given that we likely have little to no constraint on the exact shape of this feature. 

The location and amplitudes of residuals between our best fit model and the data in both space and velocity will also help us understand if we have employed an inaccurate model structure. \citet{sim15} find that large turbulence leads to spatial broadening in the azimuthal direction in the channel maps, while changes in temperature are more associated with variations in the surface brightness of the emission. Significant residuals in the azimuthal direction would suggest an incorrect assumption regarding the turbulent structure. As shown below, while there are residuals in some of the fits, none of them present the characteristic predicted for MRI turbulence in the context of the \citet{sim15} model, suggesting that our assumptions about the model structure are not substantially biasing our turbulence constraints.

\section{Results\label{results}}
\subsection{DCO$^+$\label{dco+_models}}
\begin{figure*}
\center
\includegraphics[scale=.4]{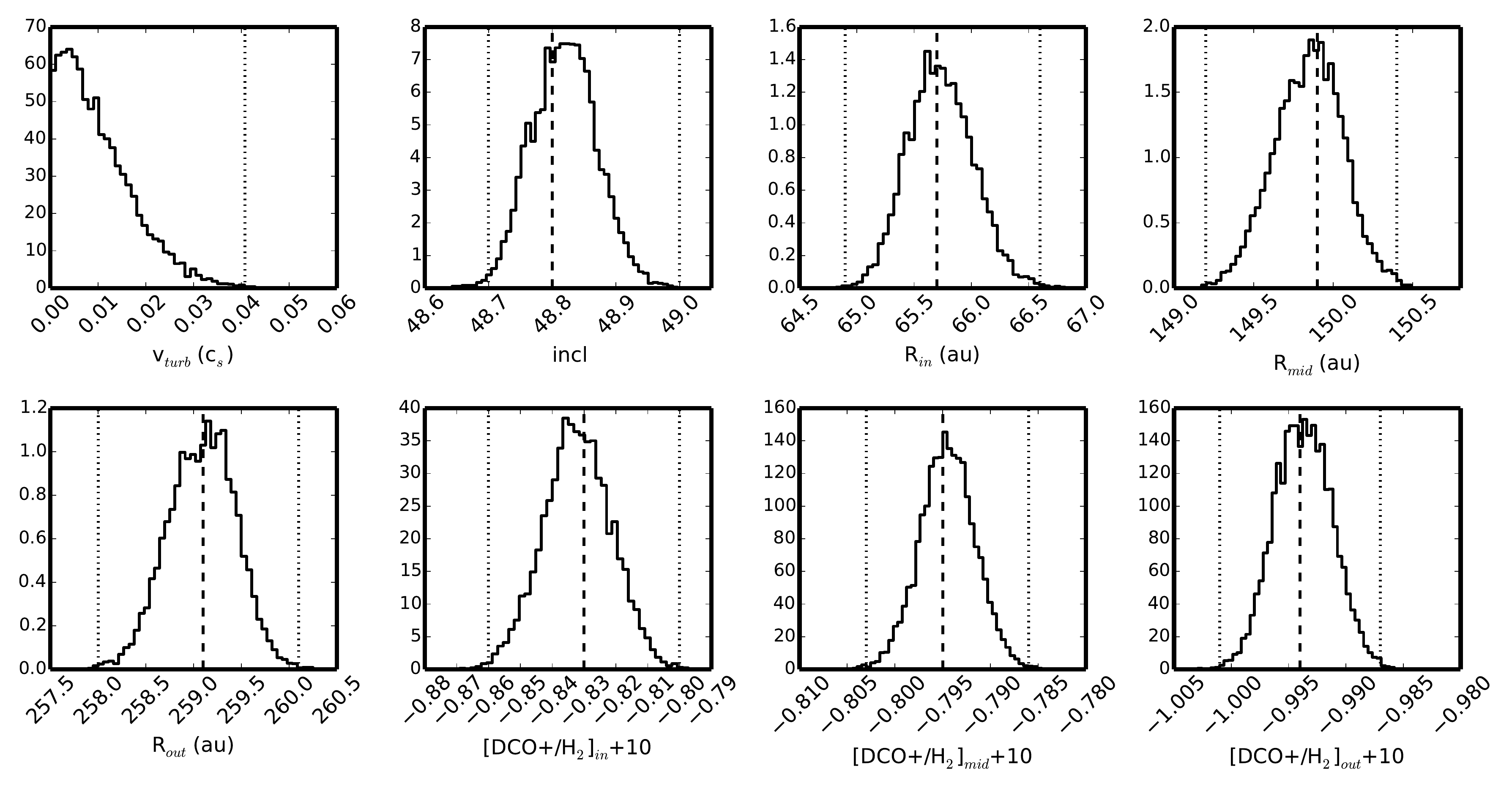}
\caption{Marginalized PDFs for the eight parameters used in fitting DCO$^+$(3-2). The shape of the turbulence PDF (top left panel) indicates that we can only place an upper limit on this parameter, ruling out non-thermal motion larger than 0.04c$_s$ at the 3$\sigma$ level.\label{pdfs_dco}}
\end{figure*}

\begin{figure*}
\center
\includegraphics[scale=.4]{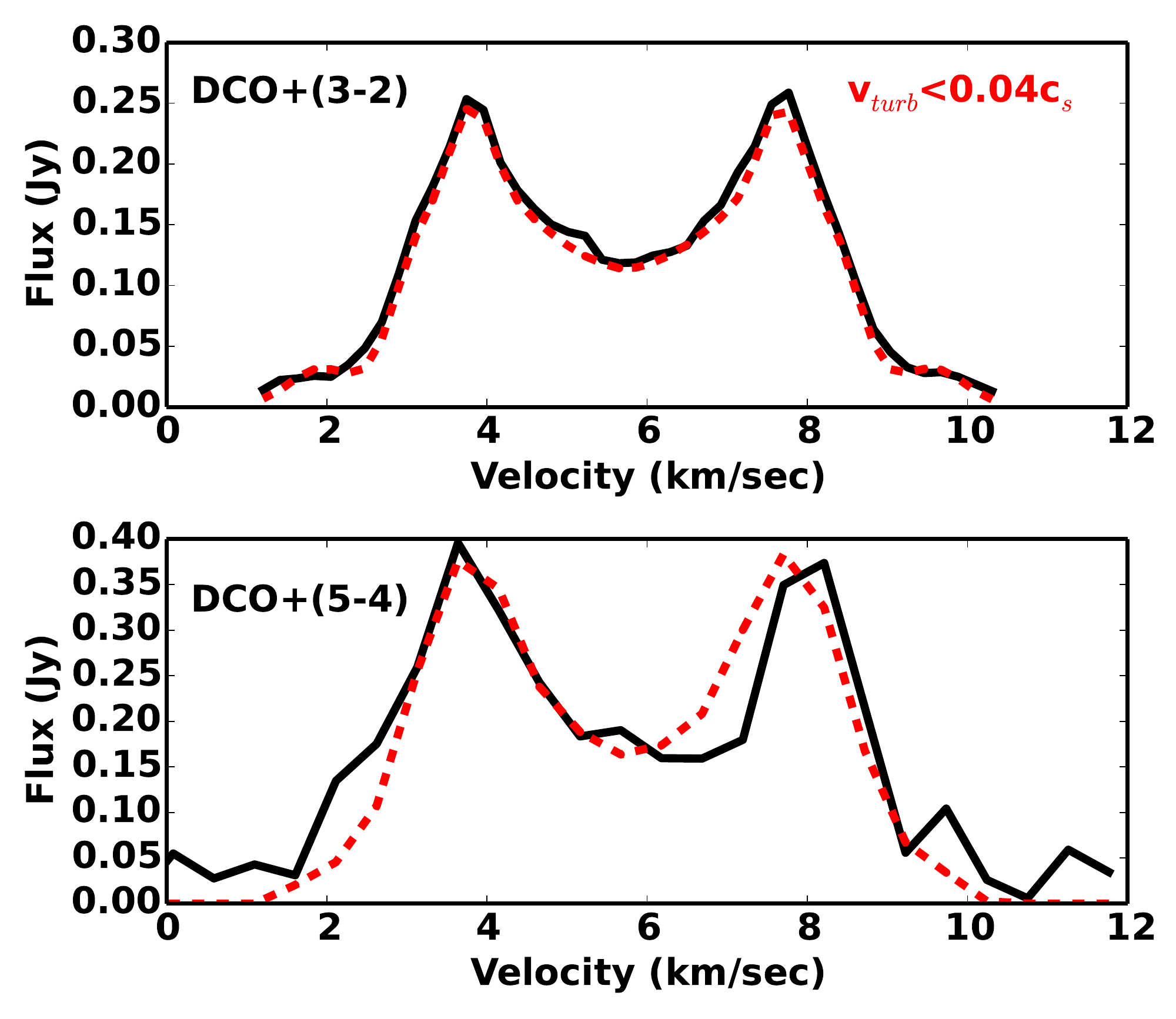}
\includegraphics[scale=.4]{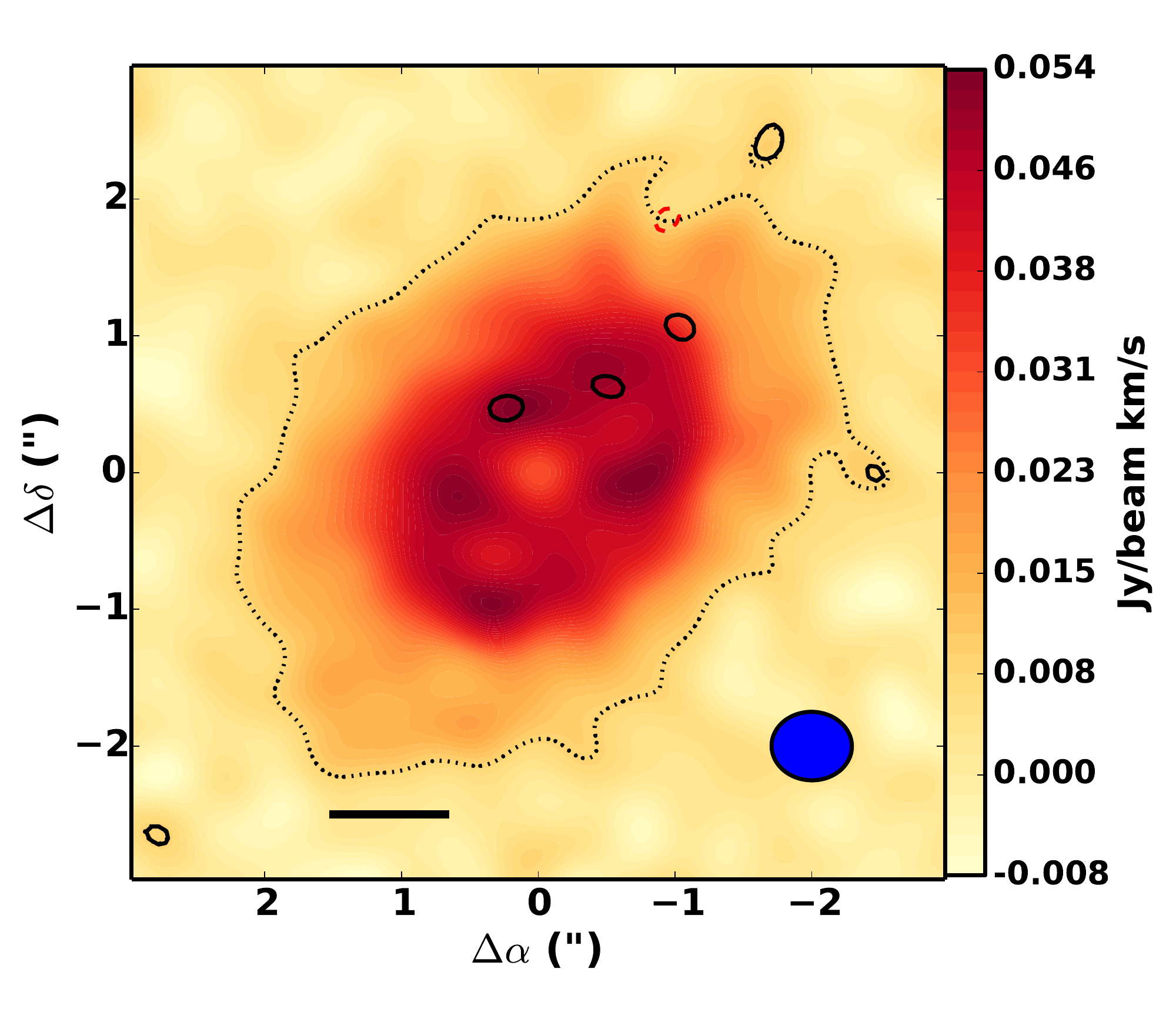}
\caption{(Left) Image spectra for DCO$^+$(3-2) and DCO$^+$(5-4) comparing the model defined by the median of the PDFs (red-dashed line) and the data (black solid line). The low turbulence model is an excellent fit to the DCO$^+$(3-2) emission, and a good match to the DCO$^+$(5-4) emission, even though DCO$^+$(5-4) was not included in the fitting process. (Right) Moment 0 map of residuals (contours) from a 3-ringed model fit to the DCO$^+$(3-2) emission (background pixel map). The dotted line marks the 3$\sigma$ ($\sigma$=0.003 Jy beam$^{-1}$ km s$^{-1}$) boundary for DCO$^+$ emission, while a 100 au scale bar and the beam are indicated in the bottom left and right corners respectively. Residual countours are shown at steps of 3$\sigma$; only small deviations are seen indicating that we can find an excellent fit without the need for strong non-thermal motion. \label{resid_dco+_moment0}}
\end{figure*}

Using our ray-tracing disk model and MCMC fitting routine, we are able to accurately confine the central radii of the three rings to 65.7$\pm$0.9 au, 149.9$^{+0.5}_{-0.7}$ au and 259$\pm$1 au (Table~\ref{dco+_results}). Posterior distribution functions (PDFs), generated using the python module corner \citep{for14}, are shown in Figure~\ref{pdfs_dco}; each parameter shows a Gaussian PDF with the exception of turbulence which is an upper limit. The model spectrum is very well matched to the data, and a moment 0 (total-intensity) map of the residuals (subtracted in the visibility domain and imaged using the clean algorithm) only shows small 3$\sigma$ features (Figure~\ref{resid_dco+_moment0}). This model is also able to match much of the DCO$^+$(5-4) emission, despite the fact that this line was not included in the fit. 

\citet{mat13} found that the DCO$^+$(5-4) emission was confined to a ring from 110 to 160 au, consistent with the middle ring in our models. \citet{qi15} fit DCO$^+$(4-3) with a ring of emission stretching from 40$^{+6}_{-3}$ au to 290$^{+6}_{-8}$ au, similar to the total radial coverage of the three rings. Our derived radial structure confirms the finding by \citet{qi15} that DCO$^+$ does not trace the CO condensation front, which is located at 90 au based on N$_2$H$^+$. The CO snow line is bracketed by the two innermost rings, with neither contributing significant emission at 90 au. 

In addition to the location of the rings, we place strong constraints on the other model parameters ($v_{\rm turb}$, inclination and abundance). We measure peak abundances within the three rings of log([DCO$^+$/H$_2$])= -10.83$\pm$0.03, -10.79$\pm$0.01 and -10.99$\pm$0.01 for the inner, middle and outer ring respectively. These values are similar to those predicted by chemical models \citep{wil07,tea15,obe15}. We are able to reproduce the data almost entirely with Keplerian and thermal motion, with any non-thermal velocity dispersion limited to $<$0.04c$_s$. In the discussion below we interpret this as a limit on the turbulence, but it also applies to other non-thermal non-Keplerian effects, such as spiral arms, disk winds or warps. These processes do not contribute significantly in the DCO$^+$ emitting region of the disk around HD 163296. 

The statistical errors on the derived parameters are small ($<$1\%) due to the large S/N of these data. The largest source of error is likely systematic effects that are not accounted for in our MCMC modeling. Here we consider three sources of uncertainty: (1) the assumption about the ring widths, (2) the uncertainty in the underlying temperature structure, (3) the 20\%\ systematic uncertainty in the amplitude calibration.

\begin{figure}
\center
\includegraphics[scale=.4]{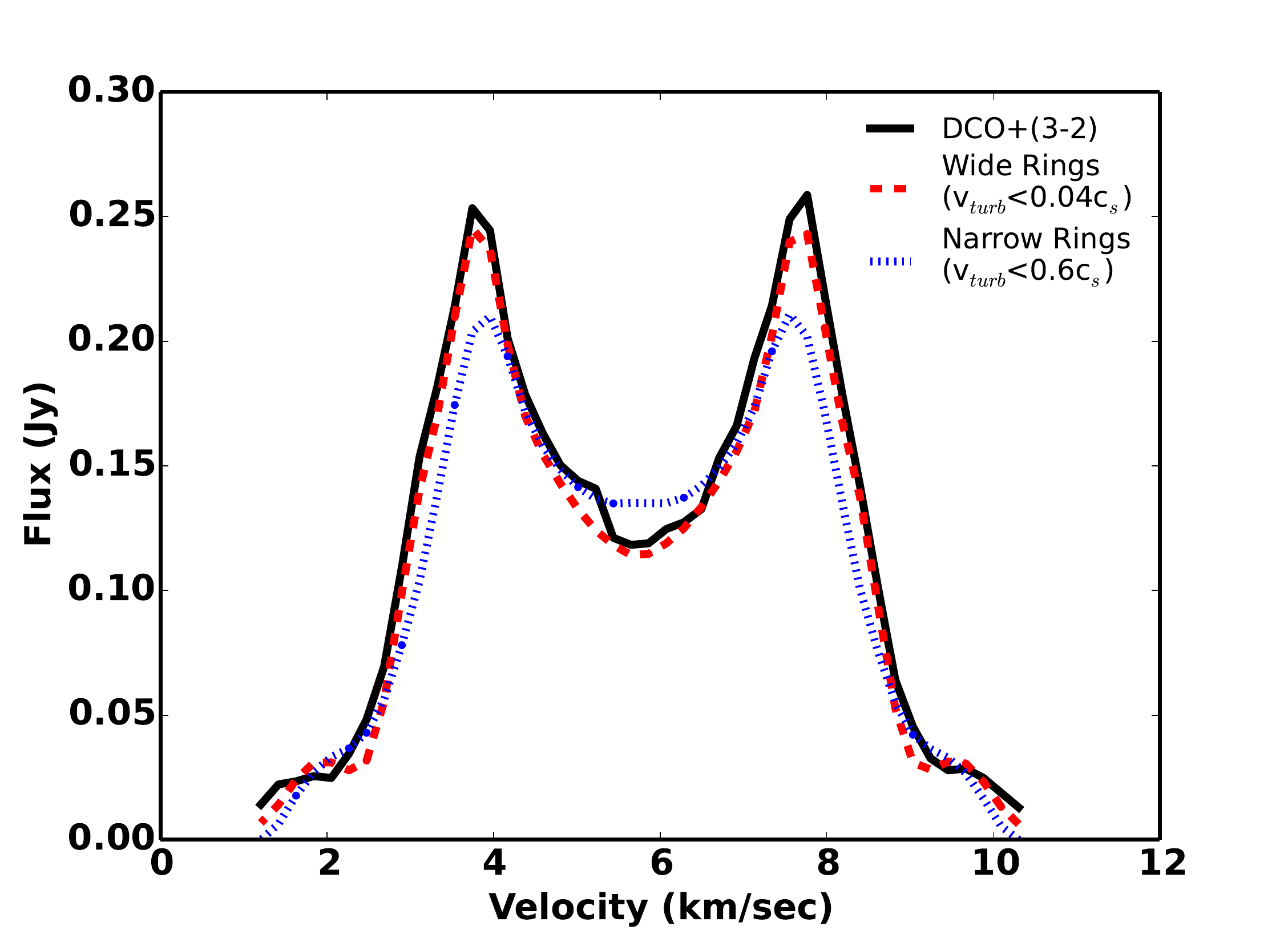}
\caption{We loom for any degeneracy between turbulence and the assumed width of the rings by re-fitting the DCO$^+$(3-2) emission with narrow radial rings. DCO$^+$(3-2) (black line) compared to a fit with rings whose width is set to 10\%\ of their radius (red-dashed line) and a fit with rings whose width is set to 1\%\ of their radius (blue-dashed line). In the narrow ring model the turbulence is much larger than the wide ring model ($<$0.6c$_s$ vs $<$0.04c$_s$) to account for the intrisically thinner features but the narrow ring model is a significantly worse fit that the wide ring model. While there is a degeneracy between the assumed ring width and turbulence, it is unlikely that we are substantially underestimating the turbulent velocity dispersion.\label{dco_narrowrings}}
\end{figure}

{\bf Narrow vs Wide Rings:} In modeling the rings, we assume each is a Gaussian with a radial width, $\sigma$, equal to 10\%\ of the rings radial location. If our assumption of $\sigma$ is incorrect, then other model parameters, such as turbulence, may be driven away from their true values in order to compensate for the incorrect assumption. To test the possibility of much narrower rings, we rerun our MCMC trial with the widths set at 1\%\ of the rings radius. We find that the fit is significantly worse, a $>10\sigma$ difference according to the Akaike Information Criterion \citep{aka74}, with the narrow ring model severely underpredicting the emission at line peak, and overpredicting the emission at line center (Figure~\ref{dco_narrowrings}). While the reduced chi-squared for these two models are similar (Table~\ref{dco+_results}), the large number of degrees of freedom in the data make this difference highly significant. It appears that the walkers have attempted to accommodate the narrow rings with increased turbulence ($v_{\rm turb}<0.6c_s$, Table~\ref{dco+_results}), but the change in turbulence cannot make up for the narrow rings. Based on these results we can assume that we have not substantially overestimated the radial width of the DCO$^+$ rings. This conclusion is consistent with the FWHM of the rings derived using the Richardson-Lucy algorithm; $\sim$40 au, $\sim$70 au, and $\sim$100 au as compared to our assumption of $\sim$20 au, $\sim$30 au, and $\sim$60 au for the inner, middle and outer rings respectively. When allowing the radial width of outer two rings to vary, we measure a FWHM of 60$\pm$3 au and 94$\pm$7 au for the middle and outer ring respectively, with a modest change in our turbulence upper limit (Table~\ref{dco+_results}). This indicates that assuming a radial width for the rings likely does not substantially bias our estimate of the turbulence. 

{\bf Temperature Uncertainty:} In our fiducial model we assumed an underlying temperature structure, but this structure is not known perfectly and its uncertainty will propagate into other quantities. In particular, the midplane temperature ($T_{\rm mid0}$) will strongly influence the derived abundances for this optically thin emission. Our knowledge of the midplane temperature is constrained in part by the CO(3-2) emission \citep[$T_{\rm mid0}$=17.5$\pm$0.25 K][]{fla15} as well as the ratio of the DCO$^+$(3-2) emission to the DCO$^+$(5-4) emission (16$\pm$4 K). To account for the uncertainty on $T_{\rm mid0}$ we run an additional MCMC trial with $T_{\rm mid0}$ as a free parameter. This trial models both DCO$^+$ emission lines, and includes a Gaussian prior on $T_{\rm mid0}$ based on the previous CO(3-2) result.

\begin{figure}
\center
\includegraphics[scale=.35]{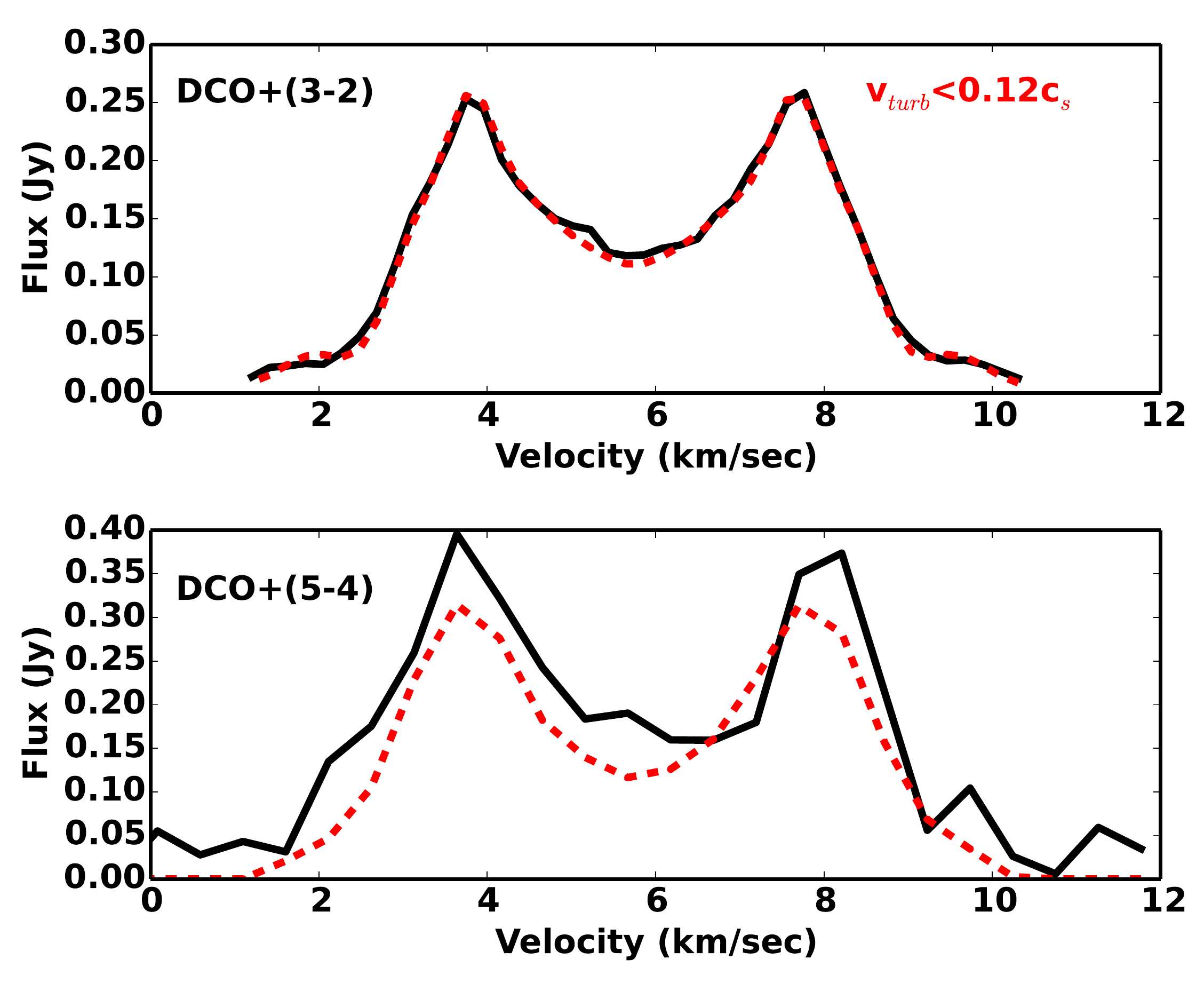}
\caption{We explore any degeneracy between turbulence and midplane temperature by fitting DCO$^+$(3-2) and DCO$^+$(5-4) simultaneously while including T$_{\rm mid0}$ as a free parameter. Spectra for DCO$^+$(3-2) and DCO$^+$(5-4) (black solid lines) compared to the median model (red-dashed line) when the midplane temperature is allowed to vary. The midplane temperature has decreased from 17.5$\pm$0.25 K to 12.2$^{+0.4}_{-0.5}$ K, with the limit on turbulence increasing to $<$0.12c$_s$. In spite of the degeneracy between midplane temperature and turbulence, we still do not detect any strong non-thermal motion in the midplane of the disk around HD 163296.\label{imspec_tmid}}
\end{figure}

\begin{figure*}
\center
\includegraphics[scale=.35]{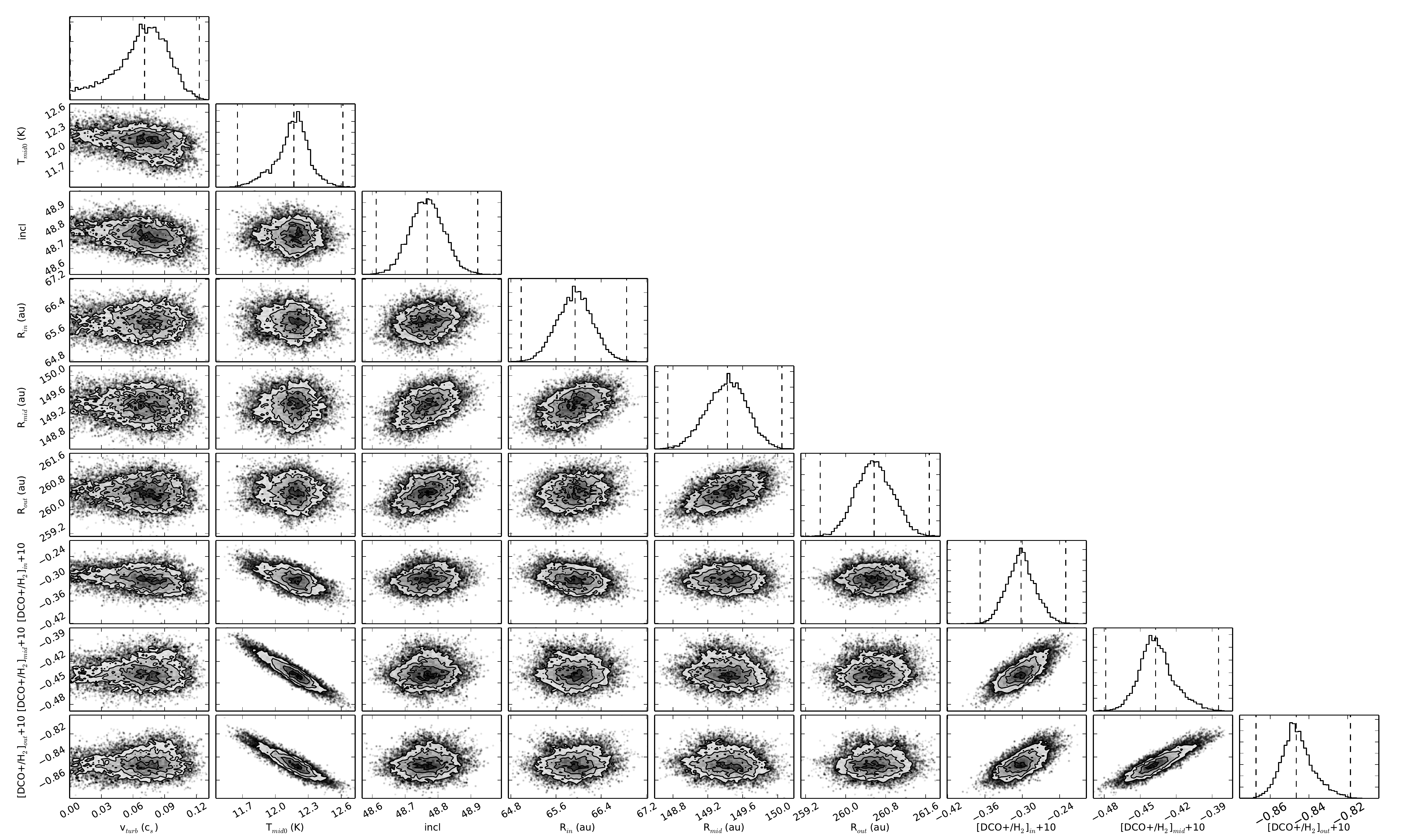}
\caption{Corner plot showing the two-dimensional posterior distribution functions, along with marginalized PDFs along the diagonal, for the model with variable $T_{\rm mid0}$ used in fitting DCO$^+$(3-2) and DCO$^+$(5-4). The shape of the turbulence PDF (top panel of left-most column) still suggests that we are only placing an upper limit on this quantity, as was found when keeping T$_{\rm mid0}$ fixed at 17.5 K. The PDFs also reveal strong degeneracies between the abundances within the three rings and the midplane temperature, as expected for optically thin emission.\label{pdfs_dco_tmid}}
\end{figure*}

Despite the prior on $T_{\rm mid0}$, we find that the walkers have converged on $T_{\rm mid0}$=12.2$^{+0.4}_{-0.5}$ K (Table~\ref{dco+_results}). With this lower temperature comes an increase in the turbulence limit to $v_{\rm turb}<$0.12c$_s$ and an increase in the rings's abundances by 0.1-0.5 dex. This lower temperature model is as effective at fitting the DCO$^+$(3-2) model as the fiducial model, and is a good match to the DCO$^+$(5-4) spectrum (Figure~\ref{imspec_tmid}). As expected there is a strong degeneracy between midplane temperature and the abundance of the three rings (Figure~\ref{pdfs_dco_tmid}), with the anti-correlation being strongest for the outer two rings. After the walkers have converged around the best fit there is no strong degeneracy between midplane temperature and turbulence, but the rise in the turbulence limit relative to the fiducial model indicates that there is some anti-correlation between the two quantities. Even when accounting for this degeneracy, we find that the turbulence PDF still shows signs that is an upper limit (with a tail that continues to zero turbulence).

The midplane temperature derived from the DCO$^+$ modeling is significantly smaller than that derived from CO(3-2) and there are a number of factors that could contribute to this discrepancy. The DCO$^+$ measurement of $T_{\rm mid0}$ relies on the relative fluxes of the J=(3-2) and J=(5-4) lines, and as a result will be strongly influenced by any uncertainty in the amplitude calibration for these data. The CO(3-2) measurement relies on the geometric separation of the near and far side of the disk, which is less severely influenced by amplitude calibration, although it is a more indirect measure of $T_{\rm mid0}$ since little to no CO(3-2) emission arises from the cold midplane. Deviations of the midplane temperature from our simple power law prescription may also influence our measurements. More detailed modeling is needed to understand the temperature structure and to reconcile the different measurements.

{\bf Systematic Uncertainty:} As discussed above, the uncertainty in the amplitude calibration of radio interferometric data is larger than the statistical uncertainties in these high S/N data sets. Here we add in the systematic uncertainty parameter using the method described above, and allow it to vary freely between 0.8 and 1.2. We find that $sys$ very quickly moves toward the upper bound of 1.2, restricting itself to a narrow range of parameter space. During this move the turbulence, inclination, and location of the three rings do not vary from the fiducial model (Table~\ref{dco+_results}), indicating that they are not strongly degenerate with the amplitude calibration. The peak abundances all increase by 0.2 dex, since the emission is mostly optically thin and its total flux is strongly correlated with abundance, assuming a constant temperature. 

To explore the low brightness case, we fix $sys$ at 0.8 and rerun the fit. We find that turbulence, inclination and the location of the three rings again do not vary from the fiducial model, while the abundances are 0.2 dex lower (Table~\ref{dco+_results}). Based on this analysis we can ascribe a 0.2 dex uncertainty on the abundance for a 20\%\ uncertainty on the flux calibration, assuming fixed temperature. 

These two trials indicate that turbulence is not strongly degenerate with the systematic uncertainty. When $sys$=1.2, where the model fits the data as though the disk is 20\%\ brighter than recorded, the limit on $v_{\rm turb}$ increases to only 0.05c$_s$. Conversely when $sys$=0.8 we recover the same limit as in the fiducial model ($v_{\rm turb}<0.04$c$_s$). This lack of degeneracy arises because a change in amplitude calibration results in a uniform change in the surface brightness of the emission in each channel, while turbulence mainly changes the total surface area of the emission in each channel \citep{sim15}. The high spatial resolution is key to breaking the degeneracy between turbulence and the amplitude calibration.

\begin{figure*}
\center
\includegraphics[scale=.3]{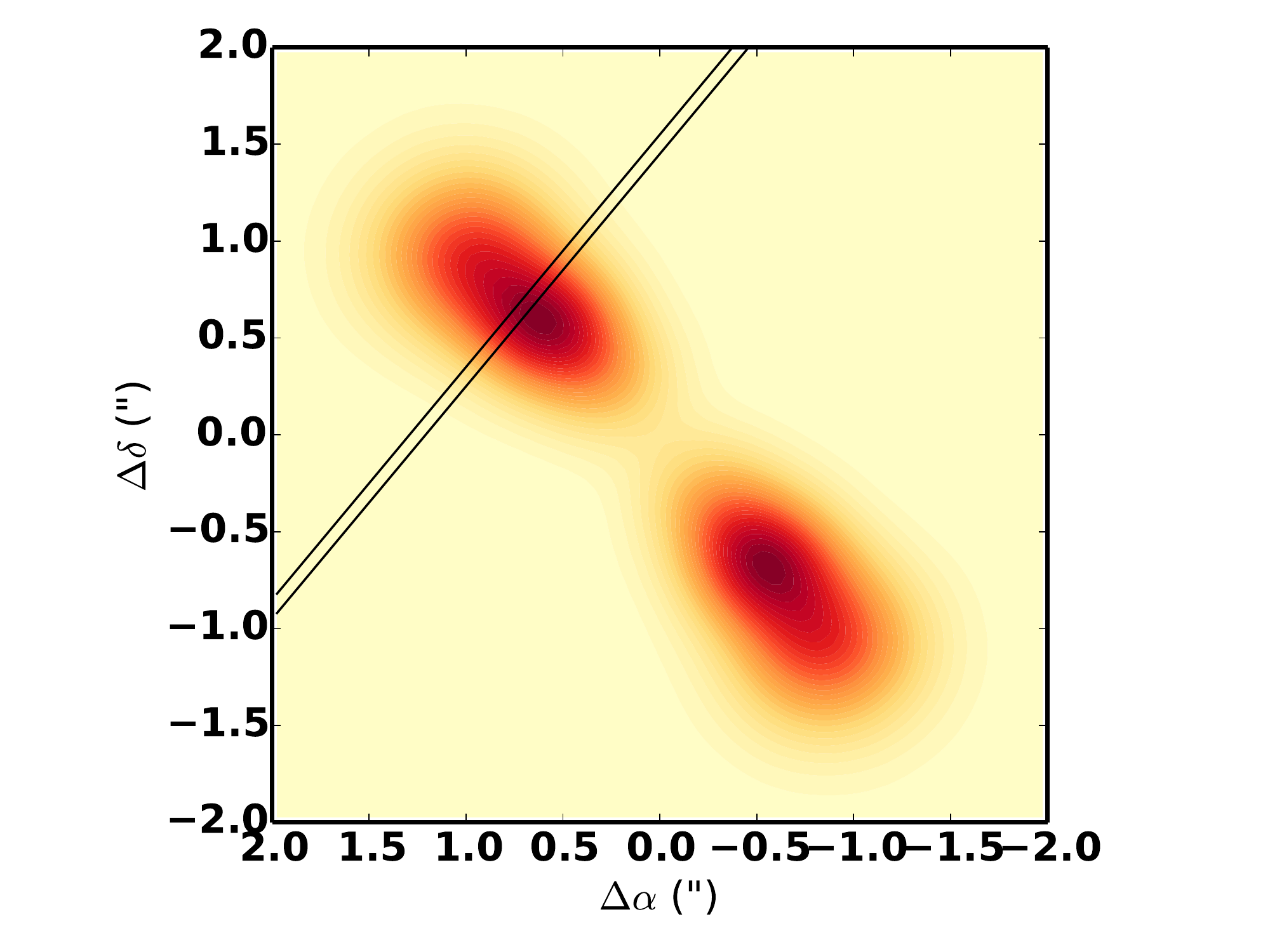}
\includegraphics[scale=.3]{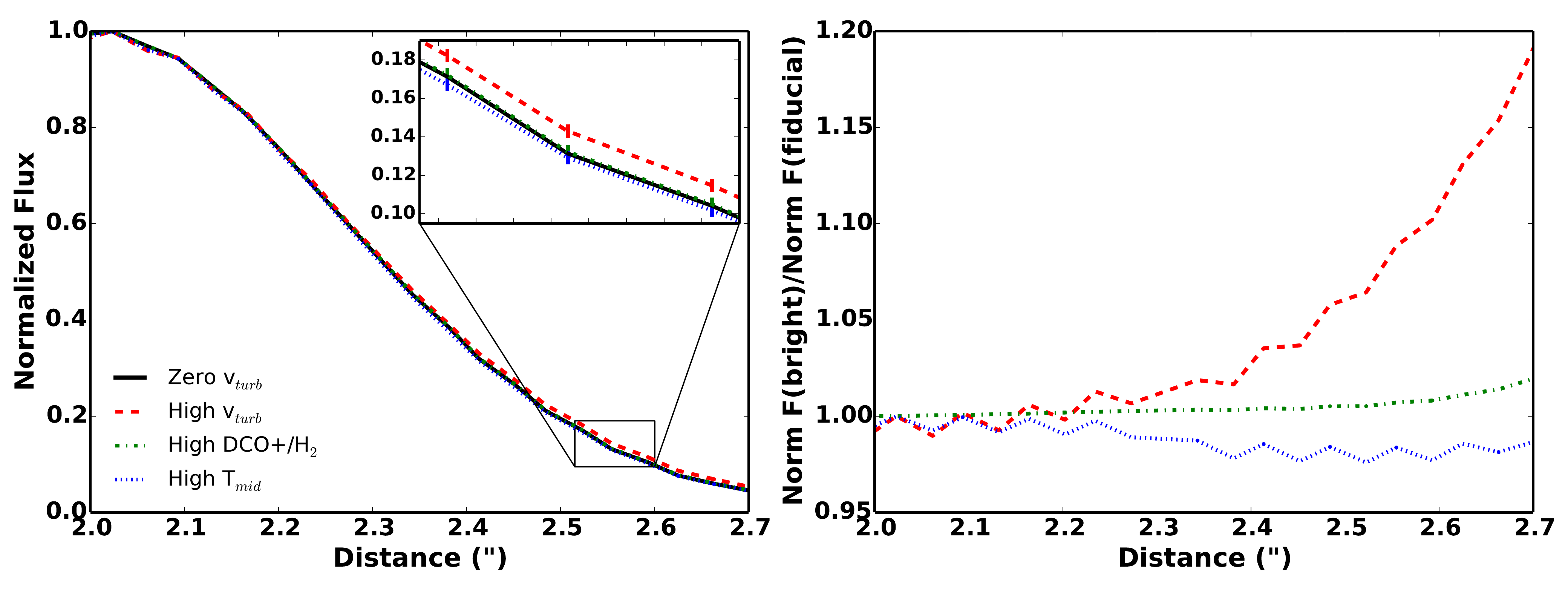}
\caption{Spatial information helps us constrain turbulence in the face of systematic uncertainty, as seen in the intensity profile along a slice perpendicular to the minor axis in the central velocity channel (marked in the top panel). Shown are the best fit model with no turbulence (black solid line) along with three models with turbulence (v$_{\rm turb}$=0.4c$_s$, red-dashed), abundance (increased by 0.2 dex, green dot-dashed) or $T_{\rm mid0}$ ($T_{\rm mid0}$=23 K, blue dotted) adjusted to increase the flux by 20\%. The left panel shows each profile normalized to its peak flux, while the right panel shows the ratio of the normalized fluxes between the bright models and the fiducial model. The uncertainty in the amplitude calibration allows for flux variations up to 20\%. This uncertainty does not contribute to our measure of turbulence since the increase in non-thermal motion needed to match such a large change in flux would generate a significant broadening of the emission profile. Because of this, turbulence is not strongly degenerate with the absolute flux calibration, allowing us to tightly constrain the midplane turbulence despite this substantial source of uncertainty.\label{slice}}
\end{figure*}

This effect is demonstrated in Figure~\ref{slice}. Here we show the intensity profile along a slice through the central velocity channel for the best fit model (with zero turbulence) and three comparison models. The comparison models have individual parameters adjusted such that the peak flux increases by 20\%; turbulence is increased to 0.4 c$_s$, DCO$^+$ abundances are increased by 0.2 dex, or the midplane temperature is increased to 23 K. Only looking at the total flux, these three models would be indistinguishable from calibration uncertainty, but the spatial information breaks some of this degeneracy. In particular, the model with high turbulence substantially broadens the intensity profile. This difference, while small, is statistically significant after integrating over the entire three-dimensional data set. Variations in the abundances do not significantly change the profile shape, explaining the strong degeneracy among these parameters and the systematic uncertainty. By modeling the entire spatial distribution of the emission we are able to leverage the fact that neighboring pixels contain complementary information; e.g. pixels at similar distances from the central star will have similar temperatures and utilizing the information from all of these pixels provides a tighter constraint on temperature than is possible by modeling a single line of sight. This approach does require assumptions to be made about the underlying structure (e.g. the midplane temperature is well described by a radial power law) but the lack of residuals suggest that these assumptions are appropriate for explaining the HD 163296 DCO$^+$ observations. 

\subsubsection{Source of the DCO$^+$ Rings}
The disk around HD 163296 is unique in having three DCO$^+$ rings and the origin of the DCO$^+$ at these locations depends on the creation of DCO$^+$ from CO. One of the main chemical pathways for generating DCO$^+$ is:
\begin{equation}
{\rm H_2D^+} + {\rm CO} \rightarrow {\rm H_2} + {\rm DCO^+}.\\
\end{equation}
The efficiency of this pathway increases as the temperature decreases, indicating that DCO$^+$ can be found in locations with cold CO. An additional warm pathway through CH$_2$D$^+$ has recently been explored by \citet{fav15}, but the low temperature derived from the DCO$^+$(5-4)/(4-3)/(3-2) line ratios and from the DCO$^+$(5-4) and (3-2) line fitting favors the low temperature pathway.

Assuming complete CO freeze-out at low temperatures, the coldest CO will be found just inside the CO condensation front. \citet{obe15} find that the inner DCO$^+$ ring in the disk around IM Lup lies just inside the CO snow line. In the disk around HD 163296 we find that the innermost ring, at 65.7$\pm$0.9 au, also lies inside the CO condensation front at 90 au \citep{qi15}.

The presence of DCO$^+$ at larger radii, and at temperatures below 19-25 K, requires a return of CO to the gas phase. One way for this to occur is due to the inward radial migration of dust \citep{cle16}. As dust grains migrate inward, the temperature of the outer disk rises due to the removal of opacity that would otherwise shield the midplane, and this increase in temperature may be enough to push the CO above its freeze-out temperature. We find that the outermost DCO$^+$ ring, at 259$\pm$1 au, lies just outside the edge of the continuum emission \citep[250 au][]{ise16}, consistent with this model. To rise above the CO freeze out temperature at this radius would require a $\sim$20\%\ increase in temperature above that predicted by our CO(2-1) model fit, consistent with the 10-30\%\ increase in temperature predicted by \citet{cle16}. \citet{hua17} also find DCO$^+$ emission beyond the dust outer edge around V4046 Sgr, similar to that observed here.

The origin of the middle DCO$^+$ ring is less clear. \citet{obe15} account for the outer ring around IM Lup using non-thermal desorption of CO by UV photons and cosmic rays penetrating to the midplane, and a similar process may be at work in the disk around HD 163296. Additional DCO$^+$ in the warm molecular layer, created through the CH$_2$D$^+$ pathway \citep{fav15}, may also contribute at these distances. The diversity of DCO$^+$ morphologies observed by \citet{hua17} suggest that between different protoplanetary disks different molecular pathways can dominate, and it is certainly possibly that different pathways dominate at different radii within a single system.

These chemical processes can generate a single broad ring covering hundreds of au \citep{wil07,tea15,obe15} and it is possible that the outer two rings are in fact one broad ring with a central depression in the middle. The depression between the two outer DCO$^+$ rings corresponds to a bright, and possibly optically thick, ring of dust \citep{ise16}. Optically thick dust will obscure our view of DCO$^+$ emission arising from the midplane, creating the appearance of a gap in an otherwise smooth distribution of DCO$^+$. Our modeling of the C$^{18}$O(2-1) and dust emission, described below, as well as the observations of \citet{ise16} does not indicate that this dust ring is optically thick, although we cannot rule out a very narrow ring of optically thick dust.  

Conversely, localized heating from a planet \citep{cle15} could lead to a localized increase in emission at the middle ring, by returning CO to the gas phase in a region where it would otherwise be frozen out onto dust grains, independent of the chemical origin of the outer ring. The middle ring is aligned with one of the dark dust rings in the high resolution observations of \citet{ise16} and if the continuum gap is caused by a large planet, localized heating by this planet could lead to an increase in DCO$^+$ emission.

All of the scenarios for explaining the outer two ring involve the return of CO from the solid phase to the gas phase before creating DCO$^+$. Recent studies have found rings of CO around TW Hya \citep{sch16} and AS 209 \citep{hua16} beyond the CO snow line, suggesting that CO desorption in the cold outer disk may be common. Searching directly for such an effect with CO is challenging given that the midplane is hidden behind the bright warm molecular layer, but there is evidence of cold CO near the midplane in the outer disk around HD 163296. \citet{qi15} find that fitting the C$^{18}$O(2-1) emission requires a CO depletion factor of only 5 in the freeze-out region. In our modeling of C$^{18}$O(2-1), discussed below, even after accounting for this small depletion factor, we find residuals at $\sim$250 au, consistent with enhanced CO abundance and/or temperature near the location of the outermost DCO$^+$ ring.

\subsection{C$^{18}$O and CO}
DCO$^+$ indicates that there is weak turbulence near the midplane. By combining this limit with those from other molecular lines whose emission arises from above the midplane we can constrain the vertical gradient of the turbulence in the outer disk. In our previous analysis of the Science Verification data \citep{fla15}, we found an upper limit on turbulence of $<$0.04c$_s$ from optically thick CO(3-2), but only $<$0.31c$_s$ from CO(2-1), that traces the kinematics of the upper edge of the molecular layer. C$^{18}$O(2-1), being more optically thin than CO(3-2) but still subject to freeze-out at the midplane, is sensitive to an intermediate layer between CO(3-2) and DCO$^+$(3-2). Using the SV ALMA data we were only able to limit turbulence to $<$0.4c$_s$ and here we use the higher S/N cycle 2 data to improve upon these limits. 

\begin{figure}
\center
\includegraphics[scale=.4]{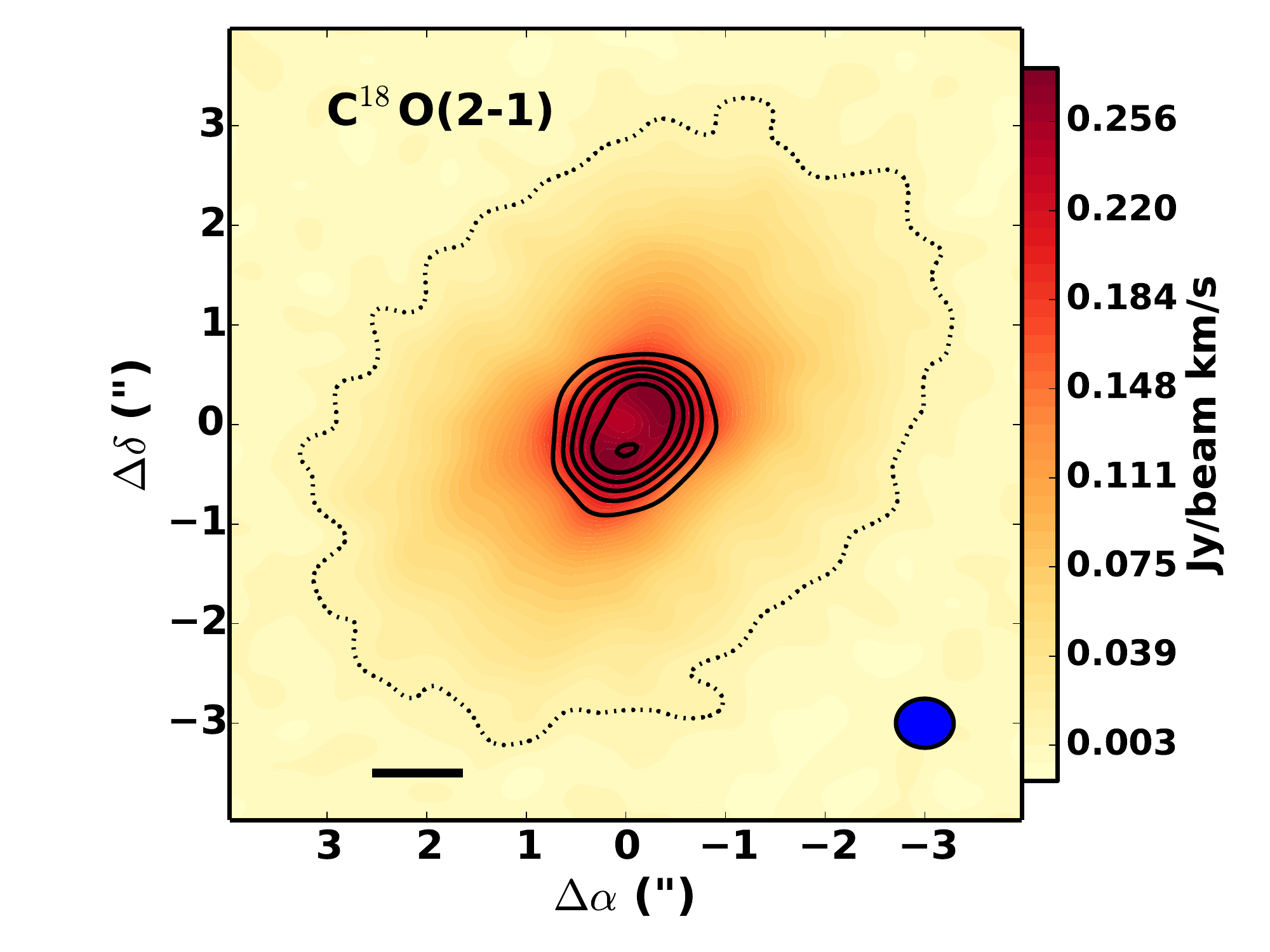}
\includegraphics[scale=.4]{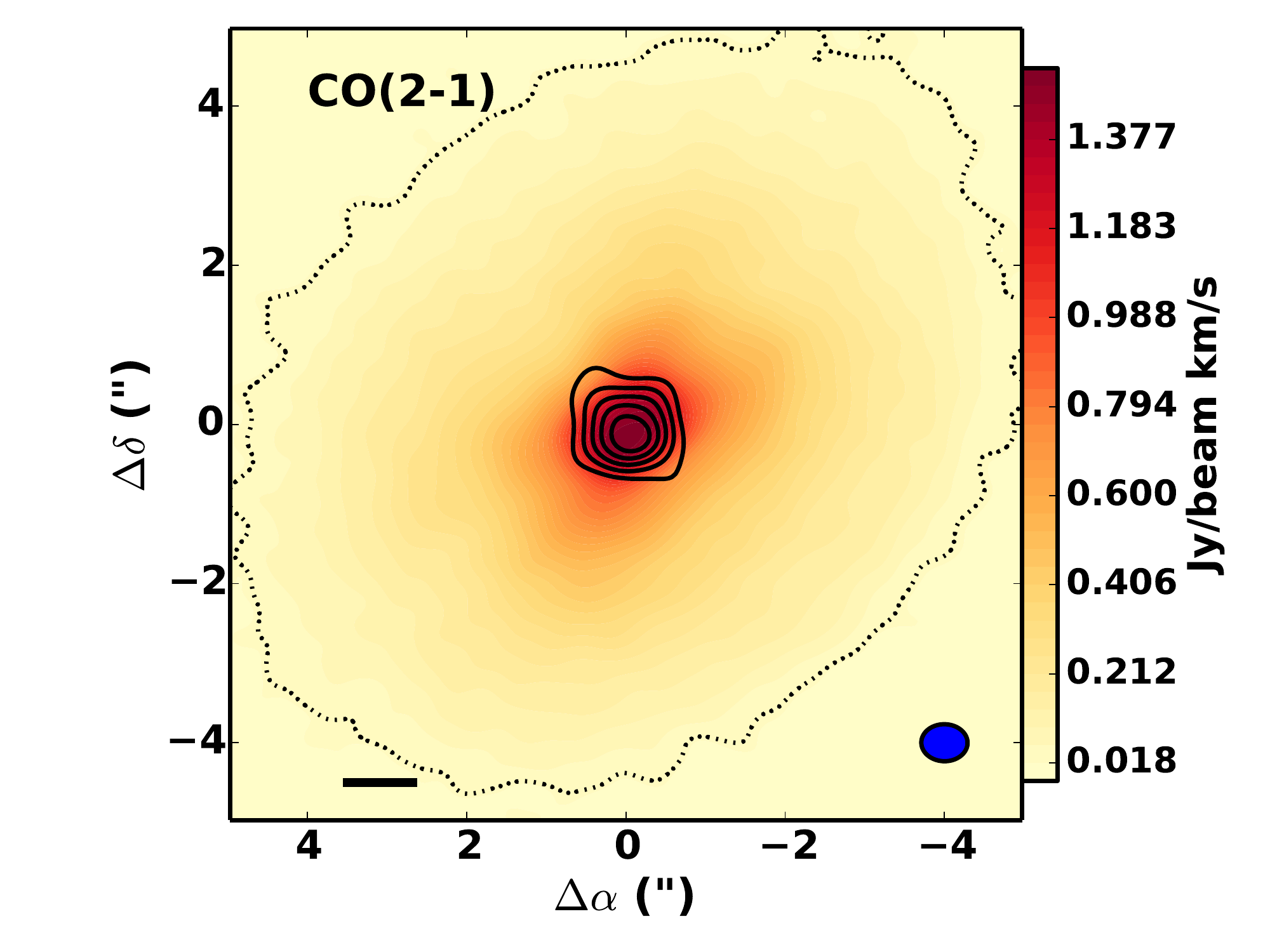}
\caption{Evidence for enhanced central emission is seen in the moment 0 map of residuals (black lines) between the best fit smooth model and the data (background image). For C$^{18}$O(2-1) the contours are in units of 5$\sigma$ ($\sigma$=3 mJy beam$^{-1}$ km s$^{-1}$) while for CO(2-1) the contours are in units of 10$\sigma$ ($\sigma$=8 mJy beam$^{-1}$ km s$^{-1}$), with the 3$\sigma$ boundary for the data marked by a dotted line in each panel. Both CO and C$^{18}$O show evidence for enhanced emission at the center of the disk, indicative of a change in the temperature and density structure from that assumed in our fiducial model. These features only represent $\sim$7\%\ of the total flux in the center of the disk, but are highly statistically significant. \label{resid_smooth_gas}}
\end{figure}

While the DCO$^+$ data is well fit with narrow rings, the C$^{18}$O and CO emission appears much smoother; image deconvolution does not reveal any prominent rings of material. We initially proceeded with fitting these data with the smooth model described above but found that the best fit models consistently underestimated the flux in the central $\sim$100 au in both lines by $\sim$7\% (Figure~\ref{resid_smooth_gas}). 

Observing an excess in both an optically thick and optically thin line suggests that it is the temperature structure that needs to be modified, rather than the CO or C$^{18}$O abundance or the surface density profile. A similar excess was observed in the gas around TW Hya \citep{ros12} possibly due to a change in temperature or a warp in the gas disk. Here we consider a change in temperature, although we cannot completely rule out other possible explanations.

The inner excess can be accommodated if $q$ changes with radius; a steeper temperature profile in the inner disk increases the temperature, and hence the emission, within this region. Such behavior is expected for an externally illuminated disk. \citet{dal06} find that the midplane temperature beyond $\sim$5 au varies as $R^{-0.25}$, while between 0.1 au and $\sim$5 au it varies as $R^{-0.75}$ (within 0.1 au viscous heating begins to play a large role). The change in temperature profile shape with radius is associated with the flaring of the surface layers, and since the midplane is heated by reradiation from the upper disk layers, its temperature profile depends on the shape of the surface layers. In the inner disk, the scale height is relatively flat, resulting in a steep drop off in temperature with radius. In the outer disk the flaring increases dramatically leading to a shallower temperature profile. While the radius for the transition from the D'Alessio models is much smaller than the size of the inner excess seen here, the model calculations were performed using a T Tauri central star, and the radial scales are expected to be larger for a more luminous Herbig star.

This feature may have been implicitly included in previous efforts to model this system. \citet{bon16} use models of an illuminated disk to reproduce C$^{18}$O observations, and find $T\propto R^{-0.5}$ out to 90 au. They also note that their models, designed to fit the C$^{18}$O(2-1) emission within 90 au, are a poor match to the outer disk flux, consistent with a change in the temperature structure between the inner and outer disk. \citet{qi15} use D'Alessio models to define the temperature and density structure, and hence include this behavior. \citet{deg13} are also able to reproduce much of the CO emission with an irradiated disk model. Our parametric models require the addition of this feature since we do not calculate the temperature structure based on the heating by the central star. 

To accommodate this behavior we explore models with a double power law temperature structure. Within $R_{\rm break}$, the radial exponent on the temperature is $q_{\rm in}$ while outside of this radius the radial exponent on the temperature is $q$. This adds two parameters ($R_{\rm break}$, $q_{\rm in}$) to our model. In modeling the DCO$^+$ emission we did not include this effect since it was not apparent in these data. Given the $\sim$100 au size scale of the inner excess, it likely only affects the innermost DCO$^+$ ring. We can estimate an approximate scale for this effect based on our DCO$^+$ modeling when T$_{\rm mid0}$ was allowed to vary. We found that the DCO$^+$ abundance of the inner ring increased by 0.5 dex when the midplane temperature decreased from 17.5 K to 12.2 K (Table~\ref{dco+_results}). In fitting C$^{18}$O(2-1) and CO(2-1) below, we find that the resulting models predict that the midplane temperature is 10-15 K higher at the location of the inner ring, which we anticipate would result in a decrease in inner ring DCO$^+$ abundance of 1-1.5 dex.

\subsubsection{C$^{18}$O(2-1)}
As noted earlier, we model the C$^{18}$O(2-1) emission by allowing $\gamma$, $v_{\rm turb}$, inclination, $X_{\rm CO}$, and $R_{\rm in}$ to vary, along with $R_{\rm break}$ and $q_{\rm in}$. We fix the remaining surface density and temperature structure parameters based on the CO(3-2) fit ($q$=-0.216, $R_c$=194 au, $T_{\rm mid0}$=17.5 K, $T_{\rm atm0}$=93.8 K). Based on the results of \citet{qi15} we assume a CO depletion of a factor of five when CO is frozen out of the gas phase, which is assumed to occur where $T<$19 K. 

\begin{figure*}
\center
\includegraphics[scale=.4]{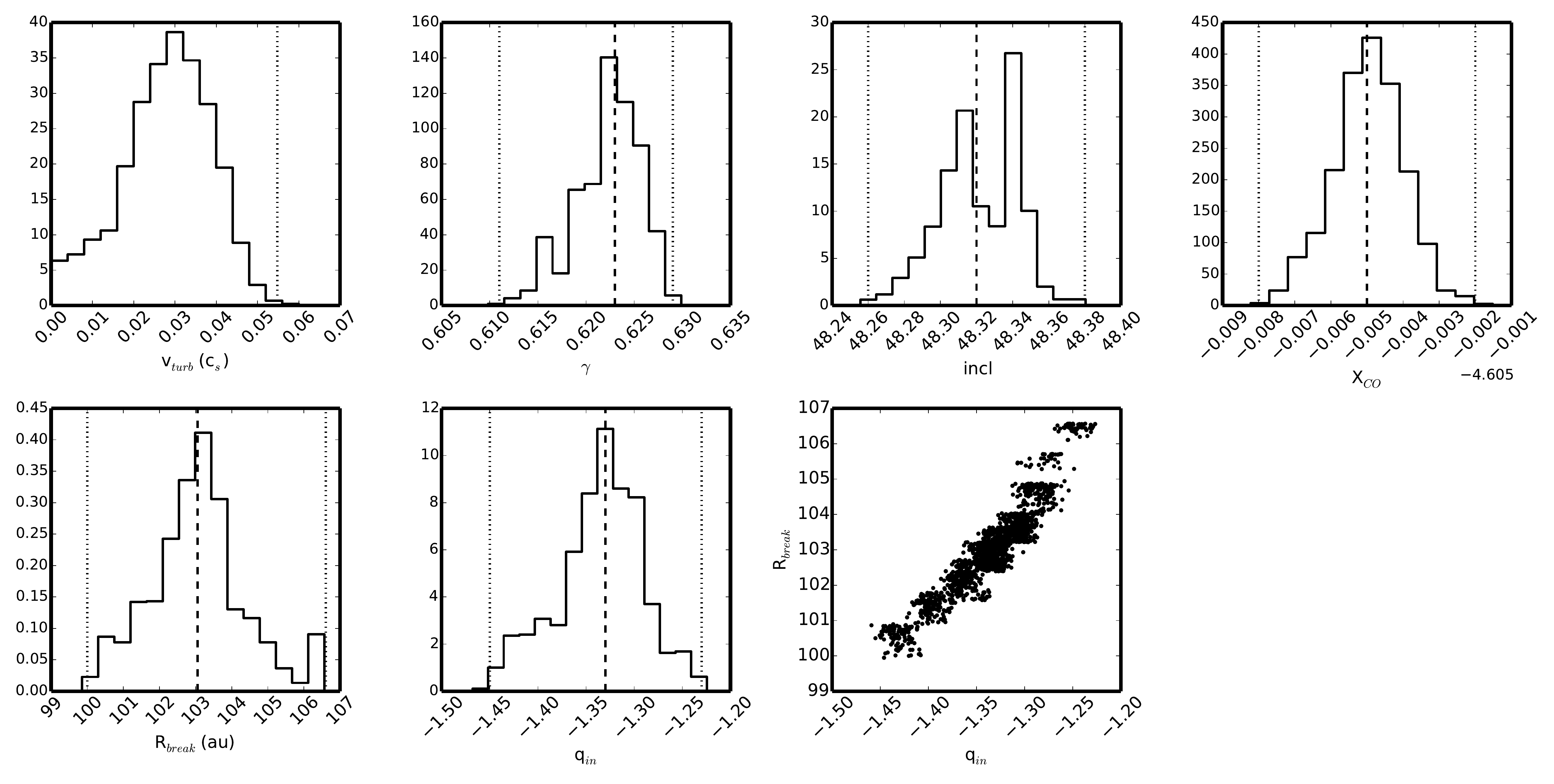}
\caption{Marginalized posterior distribution functions for the parameters used in fitting the C$^{18}$O(2-1) data. Medians are marked by dashed lines while the $\pm$3$\sigma$ ranges are marked by dotted lines. Turbulence (top left panel) is an upper limit with v$_{\rm turb}<$0.05c$_s$. The last panel shows the correlation between $R_{\rm break}$ and $q_{\rm in}$. This arises because both parameters control the flux of the inner excess, and making the inner temperature structure steeper (smaller $q_{\rm in}$) can be countered by decreasing the size of the inner excess.\label{pdfs_c18o}}
\end{figure*}

\begin{figure*}
\center
\includegraphics[scale=.4]{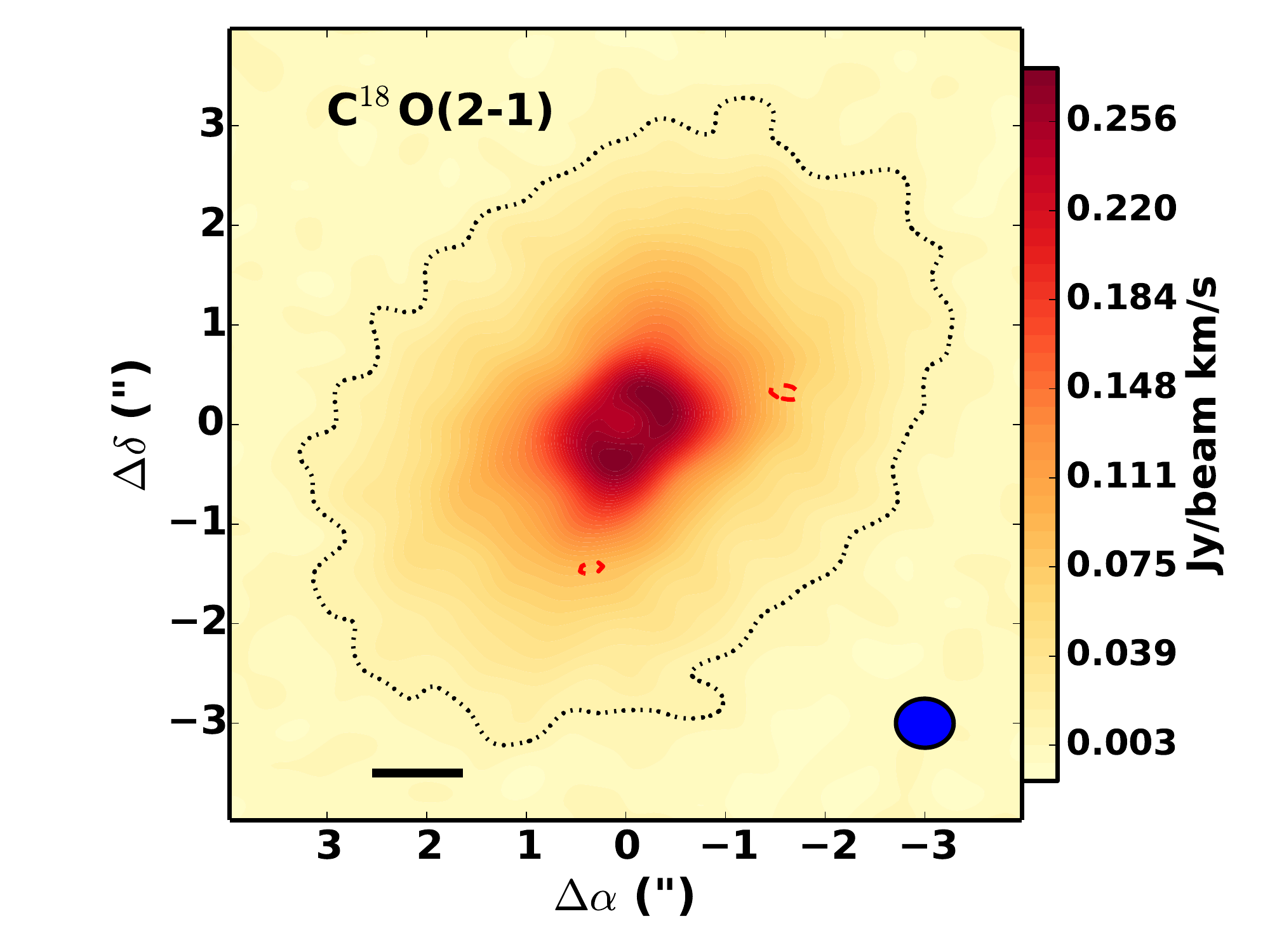}
\includegraphics[scale=.38]{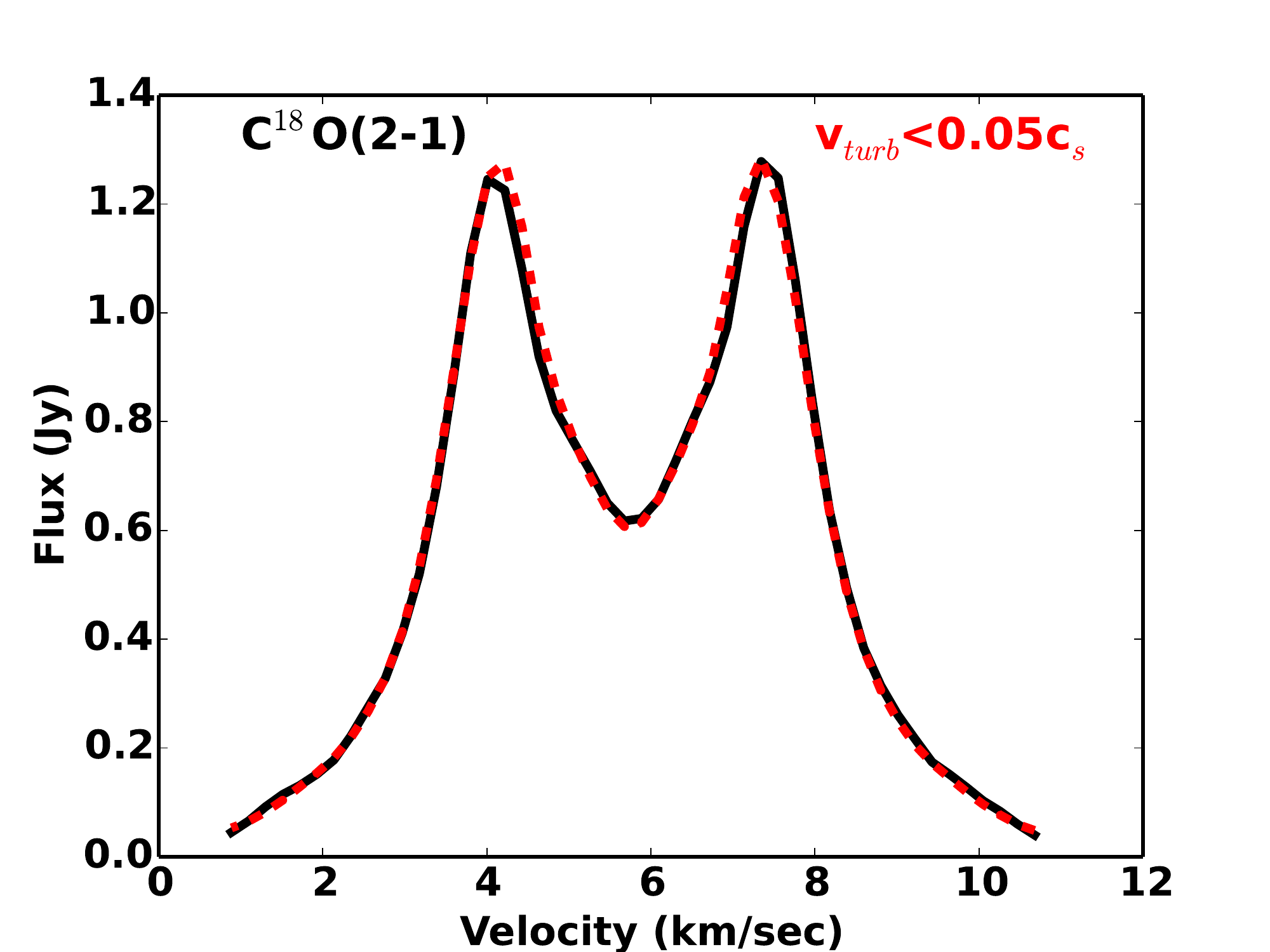}
\includegraphics[scale=.4]{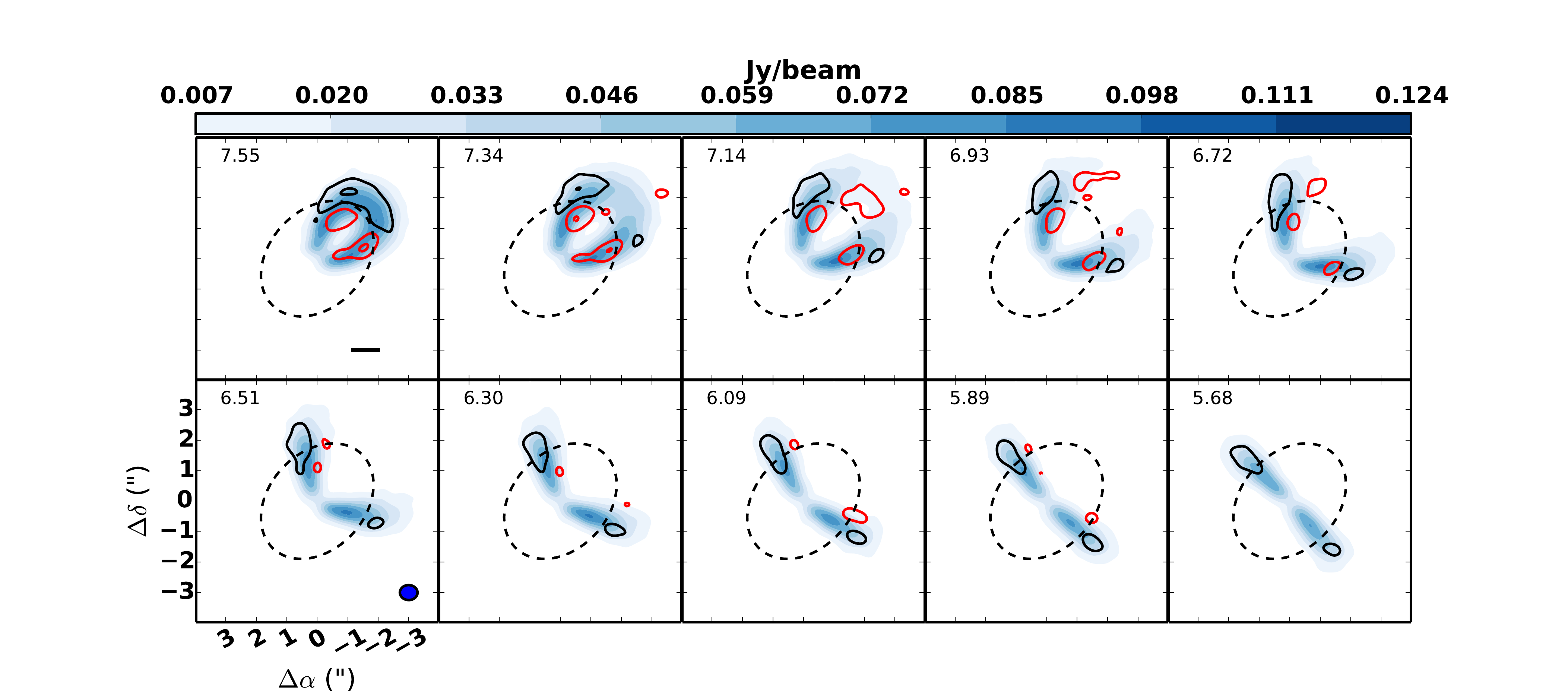}
\caption{Comparison between our best fit low turbulence (v$_{\rm turb}<$0.05c$_s$) model and the C$^{18}$O(2-1) data. (Top Left) Moment map of the residuals, with residuals marked at steps of 5$\sigma$ ($\sigma=3$mJy beam$^{-1}$ km s$^{-1}$, black contours indicate data$>$model, red-dashed contours indicate data$<$model) (Top Right) Spectrum derived from summing the flux with a 10'' box for the data (black) and model (red-dashed). (Bottom) Individual channels maps with residuals marked at 5$\sigma$, 15$\sigma$, 25$\sigma$,... ($\sigma$=1.3 mJy beam$^{-1}$). The dashed line marks a radius of 250 au. The best fit model is able to reproduce much of the emission. There are consistent positive residuals beyond 250 au, suggesting an additional reservoir of C$^{18}$O, possibly due to non-thermal or thermal desorption processes.\label{c18o_final}} 
\end{figure*}

The C$^{18}$O emission shows evidence for a resolved inner hole, which we initially constrain to $R_{\rm in}$=35.6$^{+1.7}_{-2.0}$ au. Such a tight constraint comes partly from the spatial and spectral resolution of the data, and partially from the dependence of the total flux in the inner disk on $R_{\rm in}$. Given the high S/N of this data, this latter constraint dominates the final uncertainty, but is highly dependent on the parameters defining the shape of the inner excess ($R_{\rm break}$ and q$_{\rm in}$) and the model grid resolution. For these reasons we fix $R_{\rm in}$=35.6 au for the final MCMC trial to aid in convergence of the other parameters. Final results are listed in Table~\ref{c18o_results} and marginalized PDFs are shown in Figure~\ref{pdfs_c18o}. We find that we are able to successfully reproduce much of the emission, as seen in both the spectra and the moment map of the residuals (Figure~\ref{c18o_final}). 

With this emission line we limit turbulence to $<$0.05c$_s$, similar to DCO$^+$. This limit falls below that of \citet{fla15} due to the much higher S/N, and higher spatial resolution, of these data. In the channel maps, we find positive residuals at radii $>$250 au, suggesting that we are underestimating either the CO abundance or the temperature at these large radii. \citet{cle16} find that the inward migration of dust can lead to an increase in gas temperatures of 10-30\%\ in the outer disk. The dust emission is confined to within 250 au \citep{ise16} and the increased C$^{18}$O emission beyond the dust is consistent with additional heating returning CO to the gas phase. 

\citet{qi15}, in addition to deriving a modest depletion of CO in the outer disk, also find evidence that the CO snow line occurs at a temperature of 25 K. We run an MCMC trial with with the CO snow line occurring at 25 K, instead of 19 K in the fiducial model, and find that the CO abundance has significantly increased ($X_{\rm CO}$=-4.446$^{+0.007}_{-0.008}$) and the inner temperature profile has become shallower ($q_{\rm in}$=-0.92$^{+0.09}_{-0.24}$) while the other parameters are consistent with their values from the fiducial model. In particular, turbulence ($v_{\rm turb}<0.06$c$_s$) has not substantially varied, indicating that our assumption about the location of the CO snow line does not substantially affect our turbulence measurement.

An assumption in the use of continuum-subtracted emission in the fiducial model is that the dust is optically thin. Optically thick dust can shield C$^{18}$O emission from escaping the disk, giving the appearance of a hole of gap; \citet{obe15} find an inner hole in the C$^{18}$O emission around IM Lup, which they ascribe to optically thick dust. We can characterize the influence of dust on the C$^{18}$O(2-1) radiative transfer by including dust in our models and fitting a C$^{18}$O(2-1) spectrum that has not been continuum subtraction. We parameterize the dust as a radially varying dust-to-gas ratio, with an underlying structure based on the observations of \citet{ise16}. The base dust structure is a power law in radius (dust-to-gas ratio $\sim$ R$^{-\gamma_{\rm dust}}$) that extends from 10 to 250 au, with the normalization on the dust-to-gas ratio, defined at 120 au, allowed to vary along with the radial power law exponent. We fix the three gaps at 66, 100, and 160 au with radial widths of 33, 26, and 55 au. The depletion of the dust-to-gas ratio within the gaps is treated as multiplicative constants ($dtg1$/13, $dtg1$/7, and $dtg1$/3.6, for the inner, middle, and outer ring respectively, where $dtg1$ is a free variable). This allows the model to explore an area of parameter space where the bright rings between the gaps are optically thick while the gaps themselves can still be optically thin. We also assume an opacity of $\kappa$=2.3 cm$^{2}$ g$^{-1}$ and that the dust-to-gas ratio is a constant as a function of height. We verify after the fact that the majority of the dust emission arises from close to the midplane (z/r$<$0.1) and our assumption of vertically uniform dust, without accounting for settling, will not substantially bias our results. Since the dust emission becomes most important in the inner disk, its emission will be highly degenerate with the inner excess. To alleviate this we fix $q_{\rm in}$ at -0.5, consistent with \citet{bon16}, and $R_{\rm in}$=35.6 au. 

\begin{figure}
\center
\includegraphics[scale=.4]{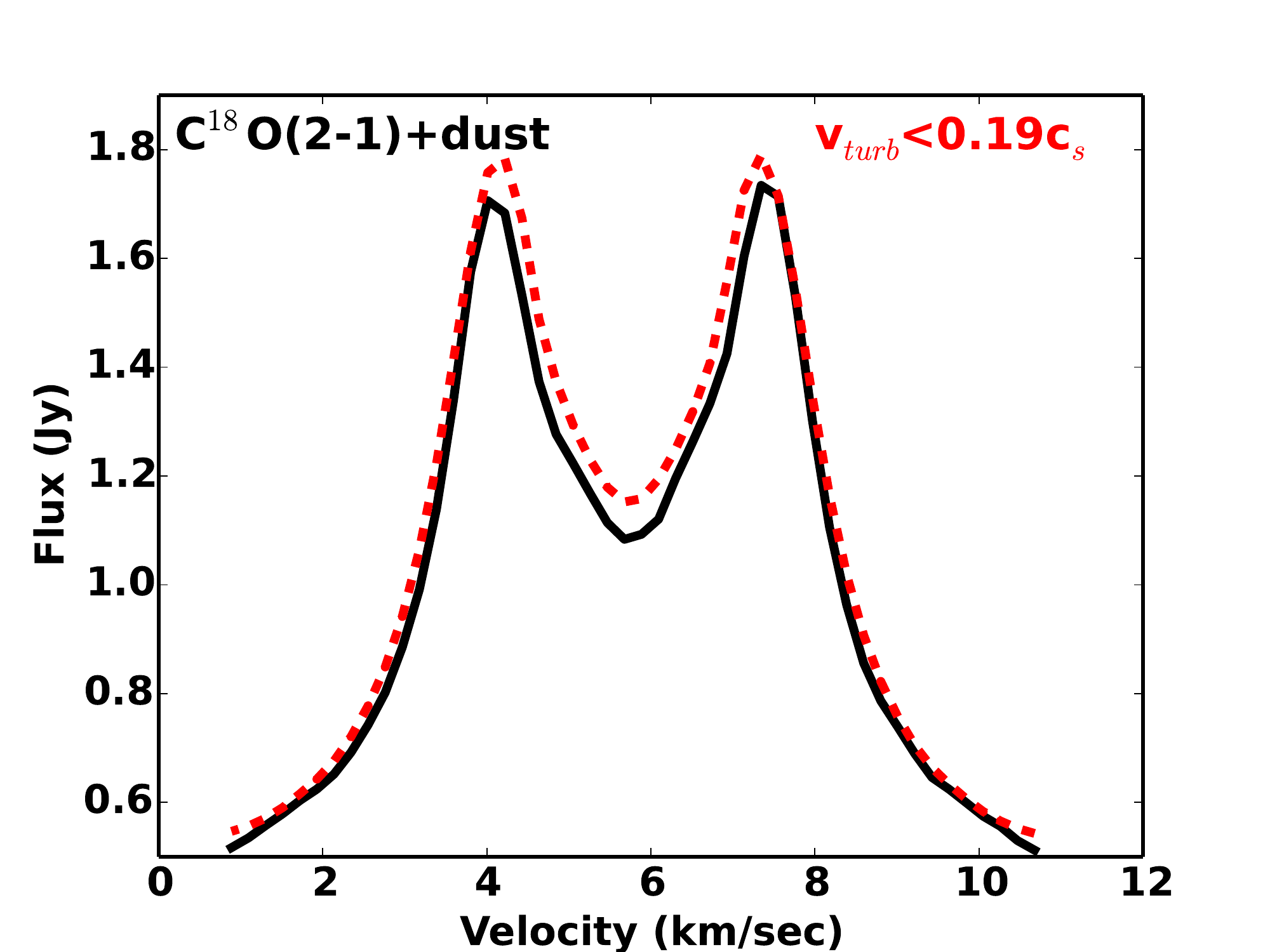}
\caption{C$^{18}$O(2-1) spectrum that has not been continuum subtracted (black line) as compared to our best fit model including dust emission (red-dashed line). We are able to obtain an adequate fit without optically thick dust, indicating that our use of a continuum-subtracted line in the fiducial model does not substantially bias our results. \label{c18o+dust}}
\end{figure}

In our resulting MCMC trial we find reasonable fits to the data (Fig~\ref{c18o+dust}) with dust that is optically thin throughout the outer disk. The dust-to-gas ratio is $\sim$0.01 with a peak optical depth of $\sim$0.5 in the inner disk and $\tau<1$ in the bright rings.  Our results are consistent with \citet{ise16} who find that the dust only becomes optically thick within $\sim$50 au, which may explain the 35.6 au C$^{18}$O inner hole but is unlikely to affect the bulk of the emission that arises from larger radii. Complete modeling of the gas and dust is beyond the scope of the paper, but the optically thin nature of the dust supports our use of a continuum-subtracted line when modeling the C$^{18}$O(2-1) emission. 

\subsubsection{CO(2-1)}
\begin{figure*}
\center
\includegraphics[scale=.4]{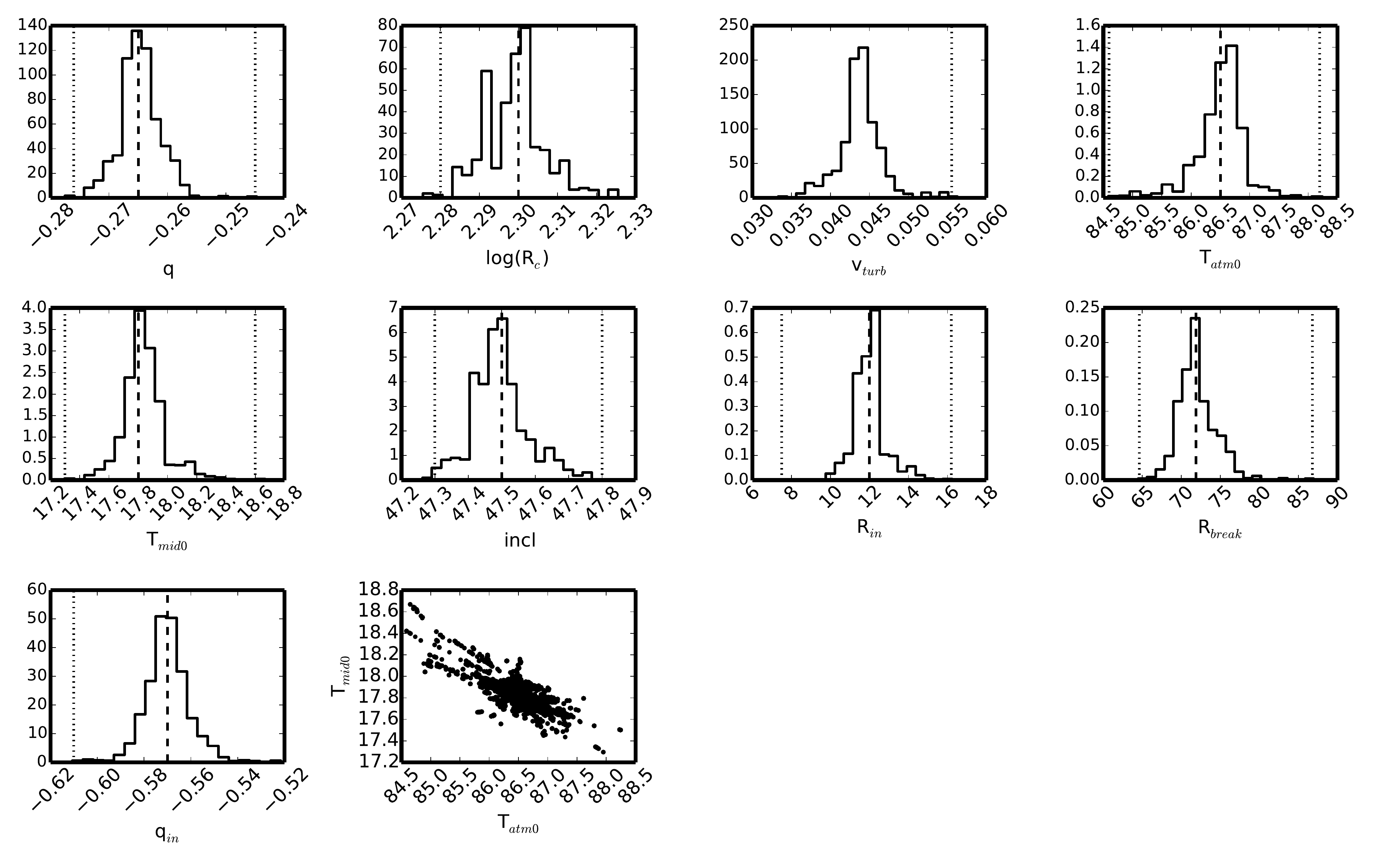}
\caption{Marginalized posterior distribution functions for the parameters used in fitting the CO(2-1) data. Medians are marked by dashed lines while the $\pm$3$\sigma$ ranges are marked by dotted lines. As with DCO$^+$(3-2) and C$^{18}$O(2-1), the PDF of turbulence rules out strong non-thermal motion (v$_{\rm turb}<$0.06c$_s$). The last panel shows the anticorrelation between $T_{\rm atm0}$ and $T_{\rm mid0}$. This arises because increasing the midplane temperature causes the disk to puff up, raising the $\tau$=1 surface to higher in the disk where the temperature is larger, and $T_{\rm atm0}$ must decrease to maintain the same flux.\label{pdfs_co}}
\end{figure*}

Similar to the CO(3-2) fitting in \citet{fla15}, we model CO(2-1) allowing $q$, $T_{\rm mid0}$, $T_{\rm atm0}$, $R_c$, inclination, and $v_{\rm turb}$ to vary, along with the addition of $R_{\rm in}$, $R_{\rm break}$ and $q_{\rm in}$. Results are listed in Table~\ref{co_results} and PDFs are shown in Figure~\ref{pdfs_co}. The inner excess extends from 11$^{+5}_{-3}$ au out to 70$^{+17}_{-5}$ au with a temperature power law slope of -0.57$^{+0.05}_{-0.04}$ in the inner disk and -0.27$^{+0.01}_{-0.02}$ in the outer disk. These slopes are similar to that predicted by radiative transfer models \citep{dal06,bon16}. We are also able to limit turbulence to $<$0.06c$_s$. 

There is a modest difference in temperature structure between the CO(2-1) fit ($T_{\rm atm0}$=87$^{+1}_{-2}$ K, $q$=-0.27$^{+0.01}_{-0.02}$) and the assumed structure in the modeling of DCO$^+$ and C$^{18}$O(2-1) ($T_{\rm atm0}$=93.8 K and $q$=-0.216). This difference in temperature structure will most significantly affect the upper reaches of the atmosphere (e.g. z$>$40 au at a radius of 100 au), which is above the emitting regions of these molecules. The difference in temperature in the CO, C$^{18}$O and DCO$^+$ emitting regions between the structure derived with CO(2-1) and that derived previously from CO(3-2) is $<$5\% and will not significantly affect our results. 

\begin{figure*}
\center
\includegraphics[scale=.26]{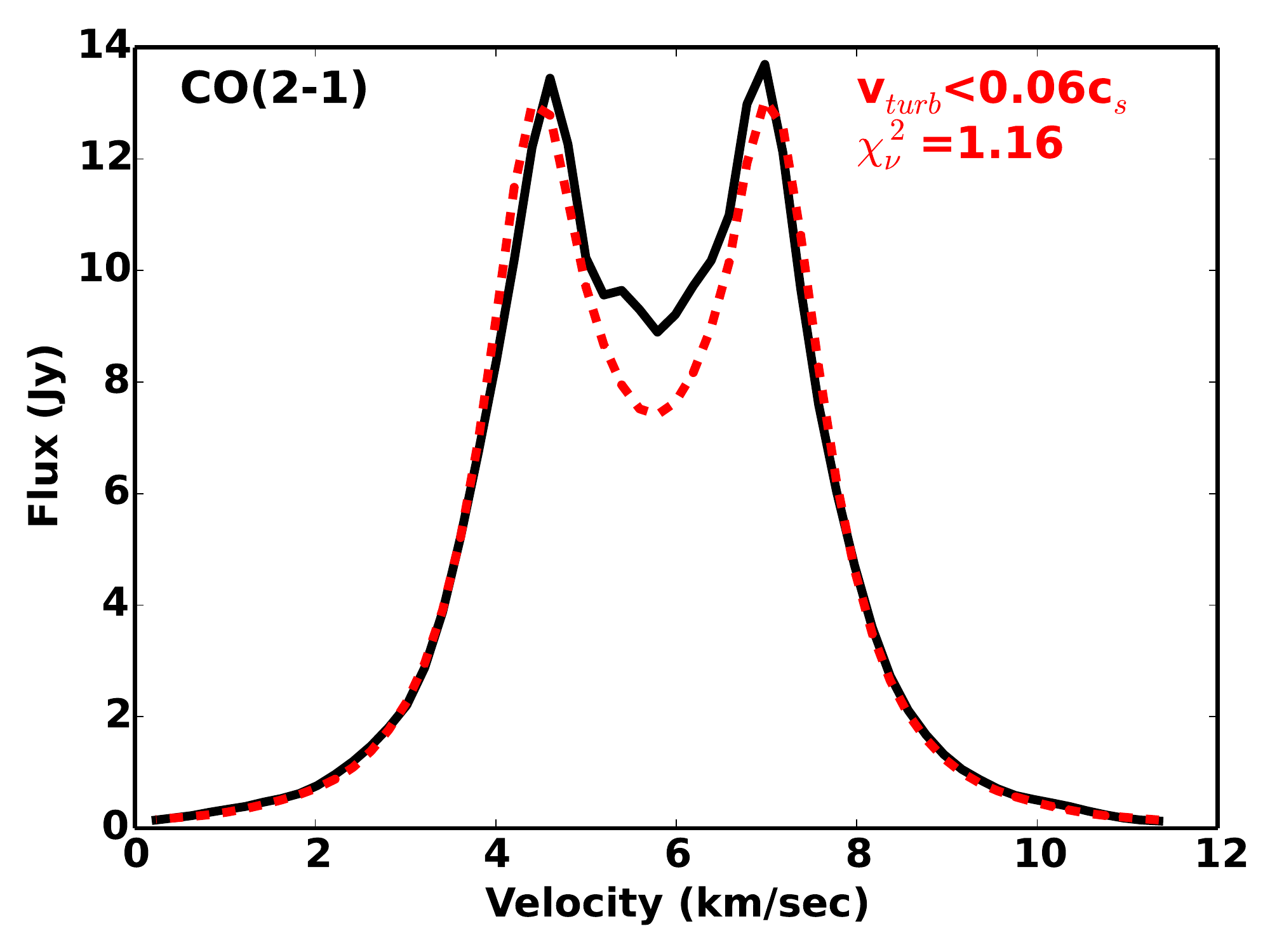}
\includegraphics[scale=.28]{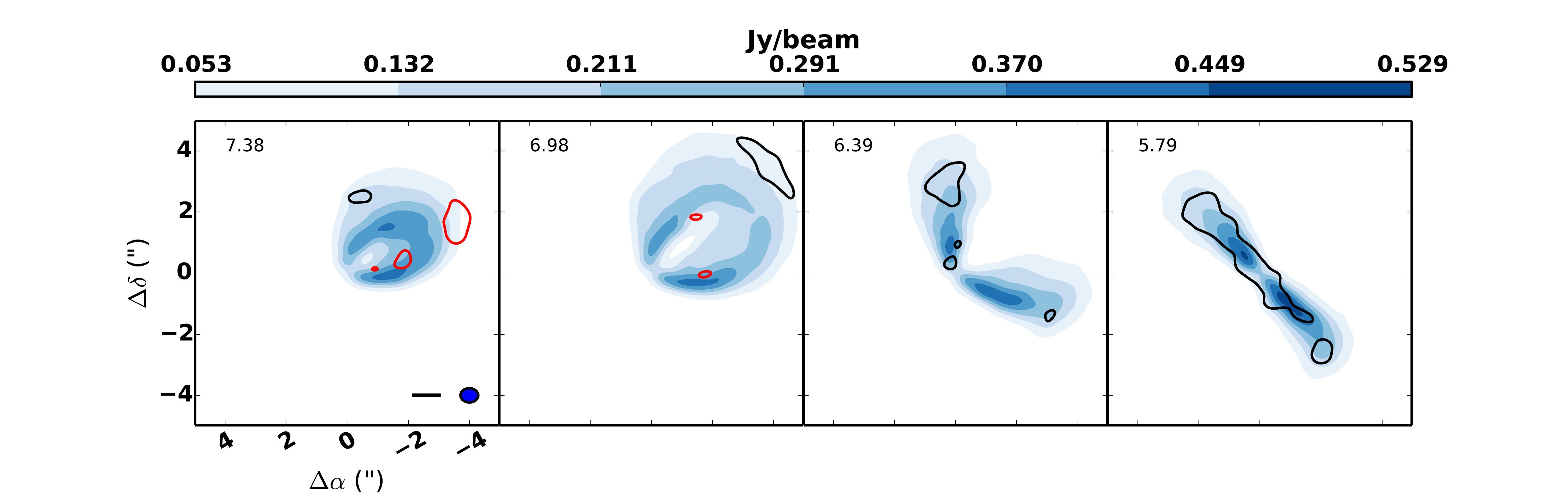}
\includegraphics[scale=.26]{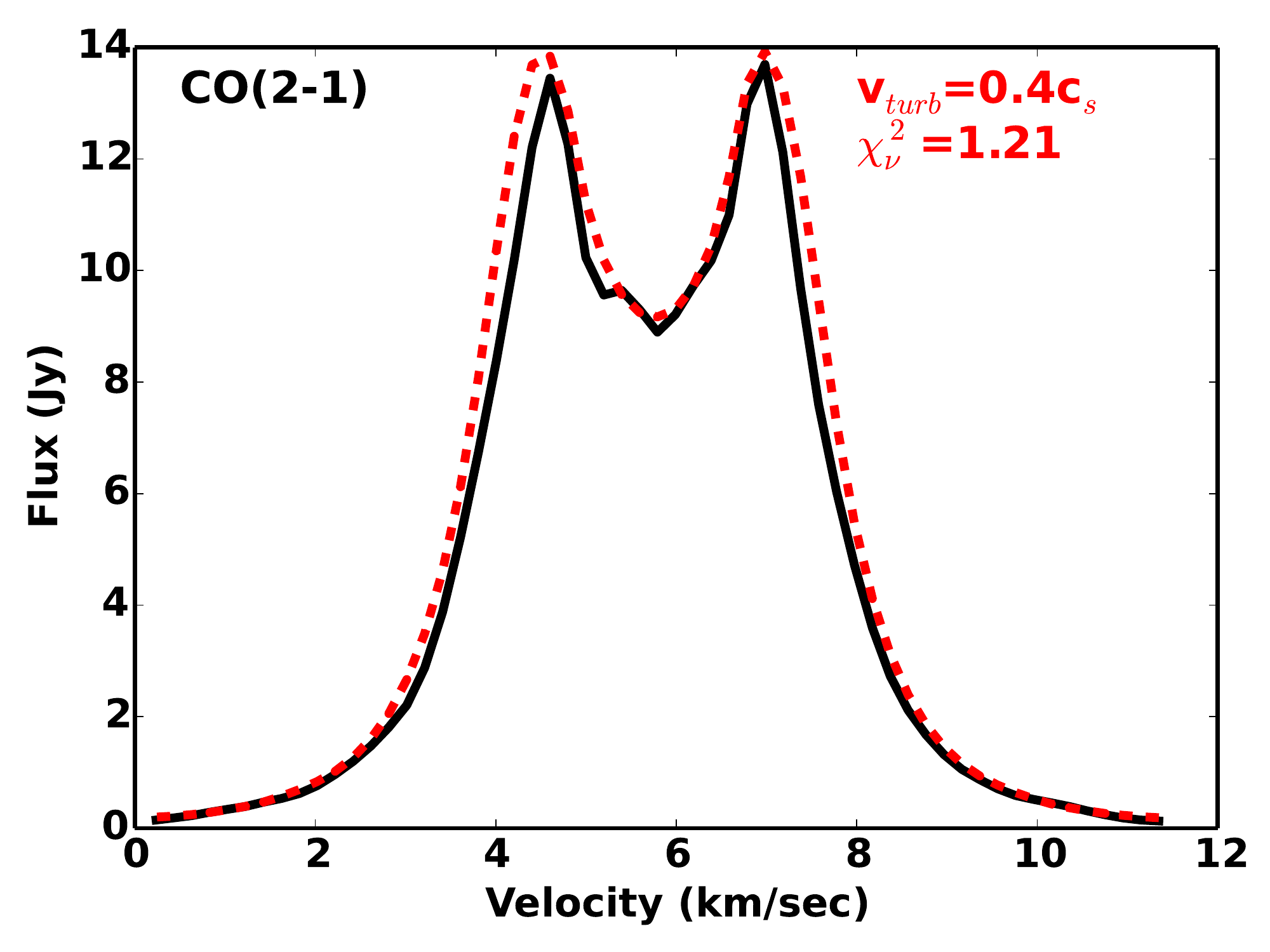}
\includegraphics[scale=.28]{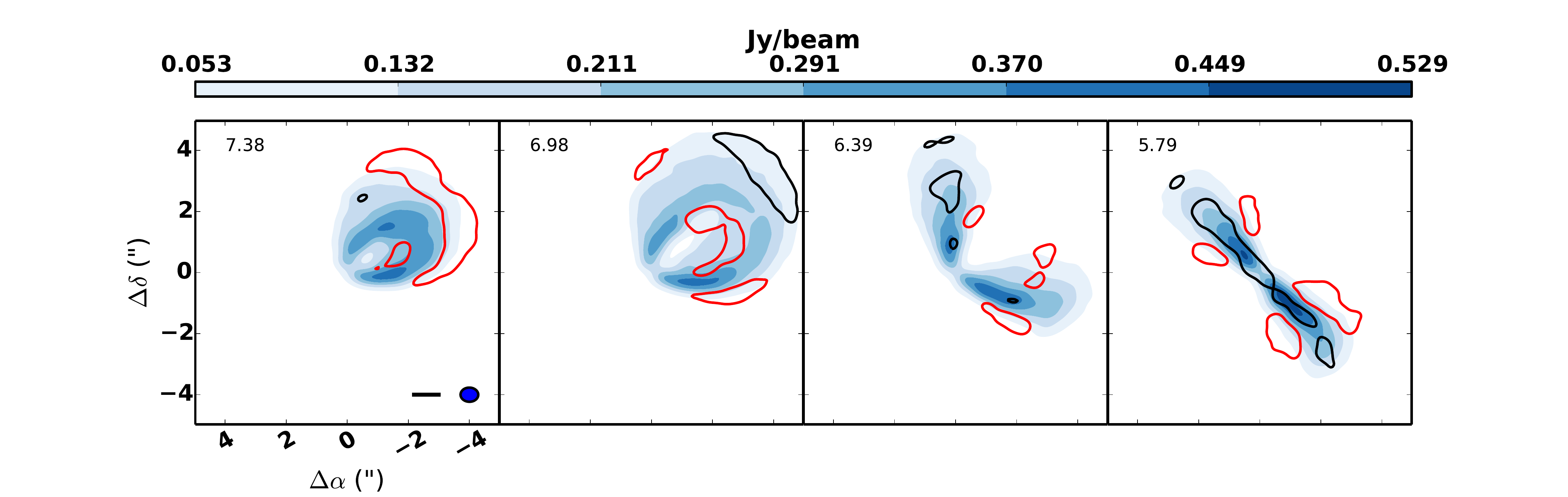}
\caption{While the low-turbulence model is not a perfect fit to the CO(2-1) data, increasing turbulence does not improve the fit. (Top Left) CO(2-1) spectrum derived from summing the flux within a 10'' box for the data (black) and low turbulence model (red-dashed). (Top Right) Individual channels maps with residuals marked at 10\%\ of the peak flux (=53 mJy beam$^{-1}$). The best fit model is able to reproduce much of the emission, with some significant positive residuals in the central channels.
(Bottom Left) Increasing the turbulence, and decreasing the $T_{\rm atm0}$ accordingly to maintain the same total flux, can produce a model that matches the spectrum, but this fit is misleading. (Bottom Right) In the channel maps it is clear that this model has not removed the residuals present in the low turbulence model, but has added negative residuals that, when summed over the entire image, cancel the positive residuals. This indicates that an increase in turbulence provides a worse global fit than the fiducial low turbulence (v$_{\rm turb}<$0.06c$_s$) CO(2-1) model.\label{co_final}}
\end{figure*}

We find a good match to the data, and the best fit has removed much of the inner excess, leaving only residuals at $\sim$3\% of the peak flux. In the channel maps we find residuals in the central velocity channels (Figure~\ref{co_final}). It appears as though our model produces emission that is more flat-topped along the azimuthal direction than the data. A similar effect can be seen in Figure 13 of \citet{sim15}, where they employ a density and temperature structure similar to that used here, but fed through the LIME radiative transfer code \citep{bri10}. This suggests that these residuals are not artifacts of the ray-tracing code, but are reflective of the assumed model structure. 

These residuals lead to a large discrepancy between the peak-to-trough ratio of the model spectrum and that of the data. \citet{sim15} found that the peak-to-trough ratio of an emission spectrum varies with the turbulence, with smaller peak-to-trough ratios associated with stronger turbulence. The large peak-to-trough ratio in the model, as compared to the data, could potentially be interpreted as a sign that the model has underestimated the turbulence, although a closer inspection reveals that this is not the case. Figure~\ref{co_final} shows a model with substantial turbulence ($v_{\rm turb}$=0.4c$_s$) that better fits the shape of the spectra. The channel maps reveal that this spectral match is misleading. The high turbulence model has not eliminated the positive residuals near line center, but has simply introduced additional negative residuals that cancel the positive residuals when integrated over the entire image. The negative residuals fall along the edge of the observed emission, indicating that the model overestimates the spatial broadening in the image plane, which is a sign of an overestimate of the turbulence. This indicates that while our model is not a perfect fit to the data, we are likely not underestimating the turbulence. 

\subsubsection{Origin of the Inner Excess}
Similar inner excess features have been observed in the gas surrounding TW Hya \citep{ros12} and the dust surrounding IM Lup \citep{cle16b}. Here we are able to model much of the emission by varying the slope of the radial temperature gradient, as predicted by radiative transfer models of flared protoplanetary disks \citep[e.g.][]{dal06}. 

The exact shape of this inner excess depends on the vertical region of the disk being traced, as well as any other parameters with a radial profile (e.g. molecular abundances, for optically thin lines) that can potentially affect the slope. Differences between the midplane temperature profile, measured by C$^{18}$O, and the atmosphere temperature profile, measured by CO(2-1), can lead to differences in $q_{\rm in}$. \citet{dal06} find that while the midplane temperature goes as R$^{-0.75}$ in the inner disk, the surface temperature profile is closer to R$^{-0.5}$. C$^{18}$O may also be subject to selective photodissociation, which can deplete its abundance in the inner disk relative to the outer disk, and relative to CO \citep{mio14}. This can explain the steeper value of $q_{\rm in}$, and larger inner hole, derived for C$^{18}$O relative to CO. Other sources of an inner excess, such as a warp \citep{ros12}, or high altitude dust grains leading to additional heating of the disk \citep{cle16b}, could potentially explain the observations. More detailed modeling that fully accounts for radiative transfer effects between different tracers is needed to constrain the combined inner and outer disk structure.

\section{Discussion\label{discussion}}
In \citet{fla15}, we used CO(3-2) to constrain the non-thermal linewidth in the upper layers of the disk to $<0.04$c$_s$. At the same time we used science verification data of C$^{18}$O(2-1) emission to constrain the motion closer to the midplane, but given the modest S/N of the data we were only able to place weak constraints on the turbulence ($<$0.4c$_s$). Here, with higher S/N C$^{18}$O(2-1) data and new DCO$^+$(3-2) data we can much more accurately characterize the turbulent motion towards the midplane of the disk. With DCO$^+$(3-2) and C$^{18}$O(2-1) we place upper limits of $<$0.04c$_s$ and $<$0.05c$_s$ respectively, while the CO(2-1) upper limit of $<$0.06c$_s$ confirms the weak turbulence in the upper layers found with CO(3-2). At the midplane in the outer disk this corresponds to $\lesssim$10 m s$^{-1}$. Under the standard $\alpha$-prescription for turbulent viscosity \citep[$\nu$=$\alpha$c$_s$H][]{sha73}, in which $\alpha\sim(v_{\rm turb}/c_s)^2$, our velocity limits imply $\alpha<$1.5$\times$10$^{-3}$, $<$2.4$\times$10$^{-3}$, and $<$3.2$\times$10$^{-3}$ for DCO$+$(3-2), C$^{18}$O(2-1) and CO(2-1) respectively. \citet{bon16}, in their modeling of the spectral energy distribution and the inner 90 au of the Science Verification C$^{18}$O(2-1) data, are able to reproduce the data with $\alpha$ of 0.1-6.3$\times$10$^{-3}$, consistent with our observations. For DCO$^+$ we have verified that our limits are robust against uncertainties in the amplitude calibration, the underlying temperature structure, and assumptions about the distribution of DCO$^+$ throughout the disk, with these effects only leading to an increase of the upper limit to $<$0.12c$_s$.  

A generic feature of MRI models of turbulence is the vertical gradient in velocity \citep{sim15,fro06,sim13b}. In these simulations, turbulence reaches 0.3-1c$_s$ at $>$3H, where H is the pressure scale height, while even in full ideal MHD simulations \citep[e.g.][]{fro06} it drops down to 0.05-0.1c$_s$ at the midplane. When accounting for ambipolar diffusion, an important non-ideal MHD effect in the outer disk regions probed by our observations, the midplane turbulence can drop as low as $\sim$0.01c$_s$ \citep{sim13b,sim15,bai15}. The disk surface, within a column density of $\Sigma$=0.01-0.1 g cm$^{-2}$, is still expected to be fully turbulent due to ionization by far ultraviolet (FUV) photons \citep{per11} leading to motions similar to those in the ideal MHD simulations, if FUV radiation is able to penetrate into the outer disk.

\begin{figure}
\center
\includegraphics[scale=.34]{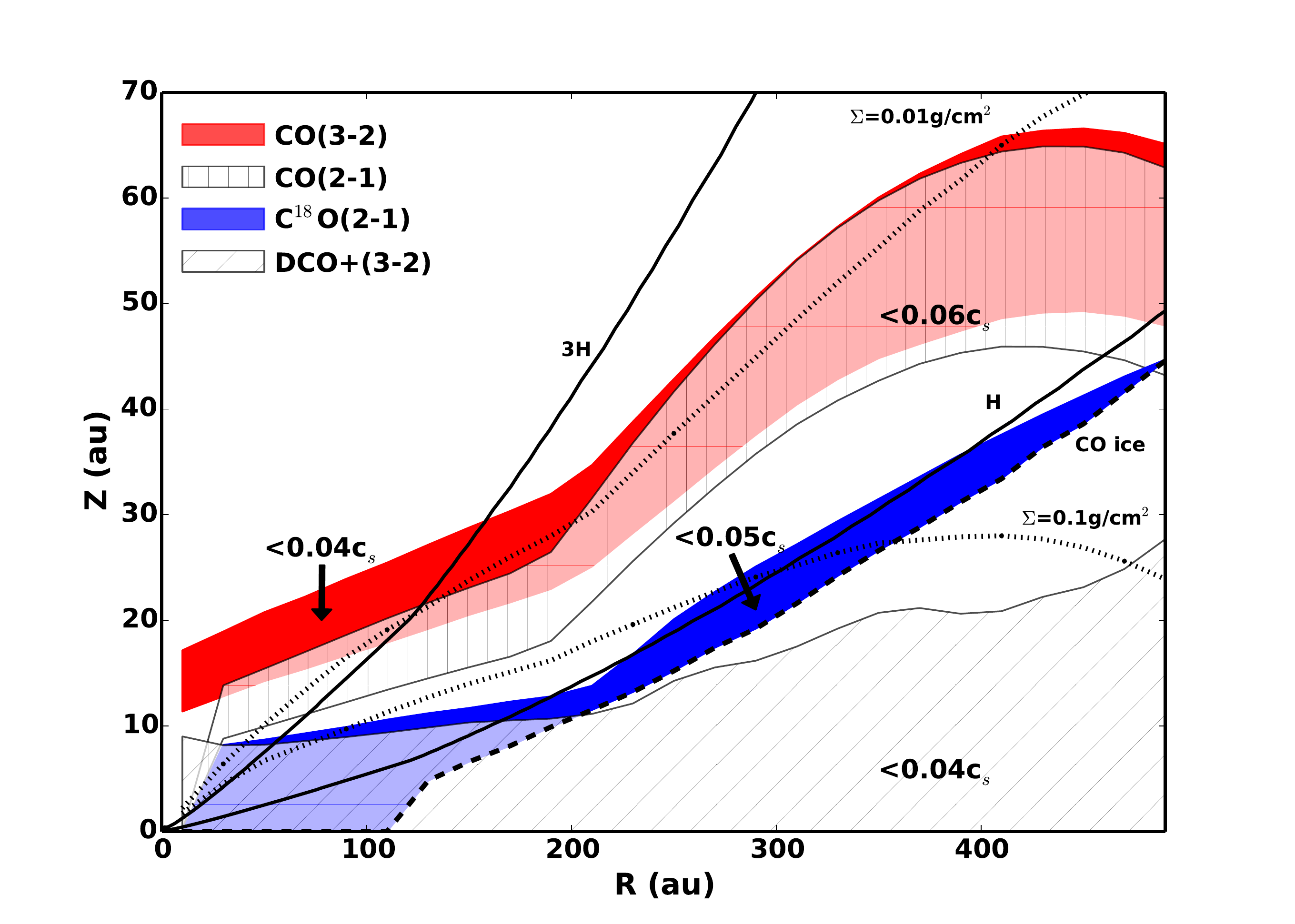}
\includegraphics[scale=.34]{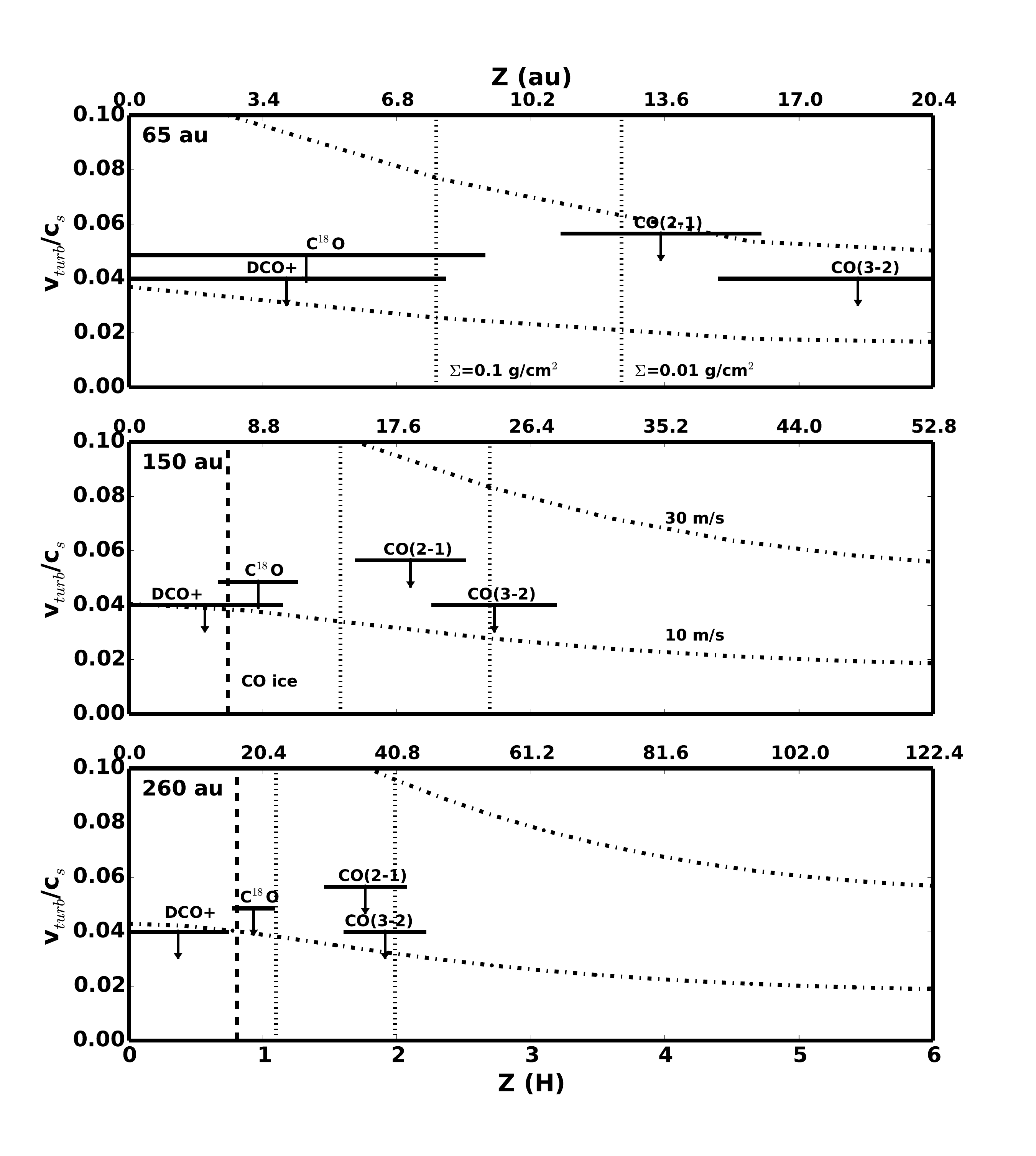}
\caption{(Left) Map of the origin of the majority of the emission from each of our molecular line tracers. Optically thick CO(3-2) and CO(2-1) arise from high up in the disk, while C$^{18}$O and DCO$^+$ trace material closer to the midplane. (Right) All four molecular lines offer complementary constraints on the turbulence throughout the vertical extent of the disk. Shown are the upper limits at the radii of the three DCO$^+$ rings. Models of MRI predict velocities of 0.3-1c$_s$ at z$>$3H and 0.04-0.1c$_s$ near the midplane \citep{sim13b}. Our upper limits for CO(3-2) and CO(2-1) fall well below these predictions in the upper layers of the disk, while our limits from DCO$^+$(3-2) and C$^{18}$O(2-1) are marginally consistent with the models near the midplane. \label{emission_map}} 
\end{figure}

To understand how our observations relate to these theoretical predictions, we first must determine the vertical origin of the CO, C$^{18}$O and DCO$^+$ emission. To do this, along each line of sight through the disk we calculate the heights above which 5\%\ and 95\%\ of the emission arises in the context of our best fit model, defining a band that encompasses 90\%\ of the line emission. We then generated an intensity-weighted average emission band as a function of radius, shown in Figure~\ref{emission_map}. The optically thick CO(3-2) and CO(2-1) lines sample the upper edge of the molecular layer with emission originating from $\sim$5H within 100 au and $\sim$2H in the outer disk, where H is defined based on the midplane temperature. Our limit on the non-thermal motion is likely more reflective of the z$\sim$2H region beyond 100 au, rather than the z$\sim$5H region within 100 au because of the larger surface area of the outer disk. At these locations in the disk, our stringent limits stand in contrast to theoretical predictions. Changes in X-ray ionization or magnetic field strength may lead to these weak non-thermal motions \citep[Simon et al. in prep]{sim13b}.

C$^{18}$O(2-1) has a much lower optical depth than CO(2-1) allowing emission from closer to the midplane to escape the disk, although it is still subject to freeze-out in the outer disk. Within 100 au C$^{18}$O(2-1) originates from below 3H while at larger radii it traces the edges of the CO condensation front. DCO$^+$, due to its low optical depth, traces the cold gas near the midplane throughout much of the disk. Our limits at these small heights are more similar to the theoretical predictions, especially when accounting for damping by ambipolar diffusion and the uncertainty in temperature on our DCO$^+$ measurement.

If the outer disk is massive enough and cool enough it may be subject to gravito-turbulence, which predicts motions of 0.2-0.4c$_s$ with little variation with height \citep{shi14}, and large radial variations only among the most massive systems that are capable of driving spiral arms \citep{for12}. The lack of strong non-thermal motion in HD 163296's disk suggest that this is not happening at a detectable level. Complex motions can also be driven by Jupiter-mass planets \citep[e.g.][]{fun16}. The recent discovery of multiple dark rings in the dust continuum \citep{ise16} hints at the presence at Saturn-mass planets. The lack of strong non-thermal, non-Keplerian motion within the gas disk is consistent with these planets being sub-Jupiter mass.

Regardless of its nature, the lack of strong turbulence has important implications for chemical mixing and dust settling. Turbulent eddies can lead to mixing of chemical species in the vertical direction \citep{sem11,fur14}. Sharp chemical boundaries, such as condensation fronts, can be softened with mixing. \citet{xu17} also find that CO gas diffuses downward into the freeze-out zone and without strong turbulent mixing the CO ice-coated dust particles are not brought up into the warmer atmosphere where they can sublimate, eventually leading to a depletion of CO from the gas phase in low-turbulence systems. \citet{fur14} find a similar result, depleted CO abundance in a low-turbulence disk, due to the build up of complex ices as a sink of carbon and oxygen in the midplane. When modeling C$^{18}$O(2-1), we find that X$_{\rm CO}$ is diminished from the standard ISM value by a factor of 4.06$\pm0.04$, consistent with these predictions. We do caution that given the degeneracies in our modeling approach the decrease in X$_{CO}$ may actually reflect a decrease in $^{18}$O abundance, dust to gas mass ratio, or a combination of all three. 

Dust grains are also affected by turbulence through their interactions with the gas. In the vertical direction, dust grain motion is subject to the competing effects of settling toward the midplane and turbulence lifting them back up into the disk atmosphere \citep{dul04}. This lifting can occur even for modest levels of turbulence, $v_{\rm turb}\sim0.1$c$_s$ \citep{fro06b}, and is especially prominent for the smallest grains \citep{fro09}. The dust settles to a scale height of h$_p$/H$\sim\sqrt(\alpha/\tau)$ when $\alpha<\tau$, where $\tau$ is the optical depth \citep{you07}. With our disk model, mm sized grains in the outer disk roughly correspond to $\tau\sim$0.1, which when combined with $\alpha<1.5\times10^{-3}$ from DCO$^+$, implies h$_p$/H$<$0.12, or h$_p$/R$<$0.007. Dust settling has been used as circumstantial evidence for $\alpha\sim$10$^{-4}$ \citep{mul12,pin16} and settling has been inferred in the disk around HD 163296 from its diminished infrared excess \citep{mee01,juh10}, although these infrared measures are more sensitive to the inner disk, while our ALMA constraints apply to the outer ($\gtrsim60$ au) disk. In modeling the C$^{18}$O and dust emission, \citet{bon16} require a level of dust settling associated with $\alpha\sim$ 0.1--6.3$\times$10$^{-3}$, consistent with our directly measured constraints. Enhanced settling associated with low turbulence may also lead to less grain growth across condensation fronts \citep{rosj13}. In general smaller $\alpha$ leads to weaker collisional velocities \citep{orm07}, and subsequently to enhanced grain growth, as has been inferred in the disk around HD 163296 based on the sub-mm spectral index \citep{nat04,ise07,gui16}. Given the lack of strong turbulence in the disk around HD 163296, radial drift may dominate over the vertical motion in setting the trajectories of the grains \citep{tur06,zhu15}. The observational evidence for radial drift in the disk around HD 163296 \citep{gui16} support the importance of this process in this system.

\section{Conclusions}
We have presented ALMA observations of DCO$^+$(3-2), CO(2-1) and C$^{18}$O(2-1) that are able to constrain the non-thermal linewidth in the disk around the Herbig star HD 163296 ($<$0.04c$_s$, $<$0.06c$_s$ and $<$0.05c$_s$ respectively). For DCO$^+$ we found that the constraint on turbulence is robust against uncertainties in the radial width of the rings and the uncertainty in the amplitude calibration, with a modest degeneracy due to uncertainty in the midplane temperature. In general, our results are relatively robust to our chose of temperature and density functional forms because each emission line traces a relatively narrow vertical region of the disk, and many different model structures can be made to pass through the constraints on temperature and density in this region. The complementary emitting regions of the three lines allow us to map the vertical structure of the turbulence and we can rule out strong turbulence from very close to the midplane up to the surface layers between radii of $\sim$30 au (the inner edge of the C$^{18}$O(2-1) emission) and $\sim$300 au (the outer edge of detectable DCO$^+$ emission). The limits on the turbulence in the upper layers of the disk in particular fall below the motions predicted by typical models of both MRI and gravito-turbulence. 

In modeling the disk, we also find that the DCO$^+$ emission is confined to three distinct rings. These rings are likely a reflection of a complex chemical structure, possibly due to non-thermal desorption of CO or the inward migration of dust grains. Bright rings of dust \citep{zha16,ise16} are not colocated with the DCO$^+$ rings, consistent with the DCO$^+$ structure arising due to complex chemical processes rather than variations in the underlying gas surface density. In the CO and C$^{18}$O emission we find an inner excess that can be explained by a change in the temperature structure towards the inner disk. These observations highlight the complex structure of planet-forming disks around young stars.

\acknowledgements
We thank the referee for their thoughtful comments that greatly improved the manuscript. KF and AMH are supported by NSF grant AST-1412647. SR was supported by the Keck Northeast Astronomy Consortium REU program through NSF grant AST-1005024. JBS and PJA are supported by NSF grant AST-1313021. We thank I. Cleeves, J. Huang, R. Teague, and A. Isella for valuable discussions. This paper makes use of the following ALMA data: ADS/JAO.ALMA\#2013.1.00366.S and ADS/JAO.ALMA\#2011.0.00010.SV. ALMA is a partnership of ESO (representing its member states), NSF (USA) and NINS (Japan), together with NRC (Canada), NSC and ASIAA (Taiwan), and KASI (Republic of Korea), in cooperation with the Republic of Chile. The ALMA Observatory is operated by ESO, AUI/NRAO and NAOJ. The National Radio Astronomy Observatory is a facility of the National Science Foundation operated under cooperative agreement by Associated Universities, Inc. The research made use of Astropy, a community-developed core Python package for Astronomy \citep{astropy13}. We thank Wesleyan University for time on its high performance computing cluster supported by the NSF under grant number CNS-0619508. This work was supported by the Momentum grant of the MTA CSFK Lend{\"u}let Disk Research Group.

\appendix
\section{MCMC Walkers}
\begin{figure}
\center
\includegraphics[scale=.35]{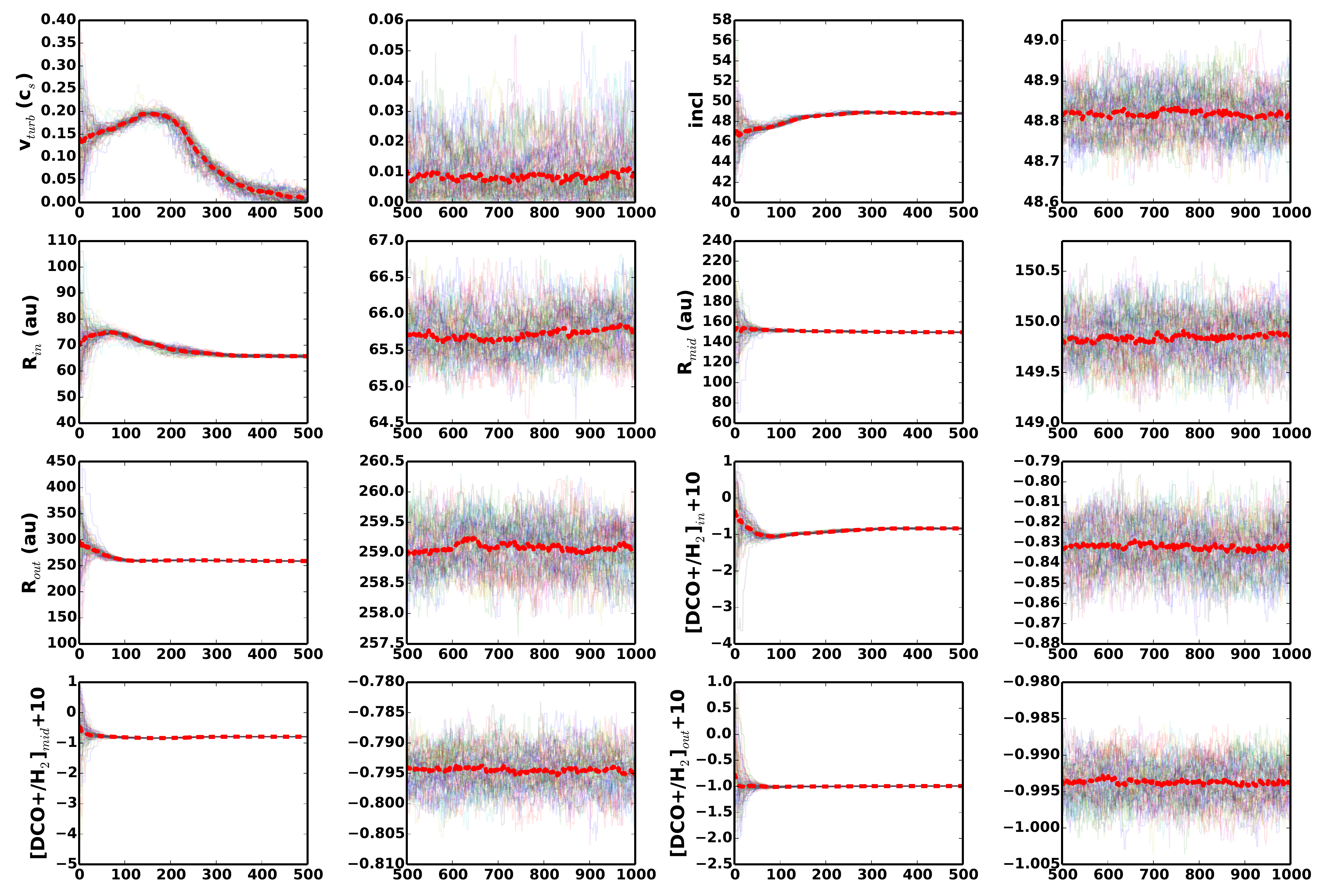}
\caption{ MCMC chains for the 80 walkers over 1000 steps in the fiducial DCO$^+$(3-2) model fit. The first and third columns show the first 500 steps, while the 2nd and 4th columns show the last 500 steps. Individual lines show the movement of individual walkers, while the red-dashed line marks the median of the walker positions at each step. The walkers are initially spread over a large region of parameter space, but quickly converge toward the best fit, with the first 500 steps thrown out as burn-in. \label{dco+_chains}}
\end{figure}

Figure~\ref{dco+_chains} shows the progress of the chains for our DCO$^+$(3-2) fiducial model fit, described in section~\ref{dco+_models}. The walkers quickly converge and settle around the best fit by step 500, with most parameters reaching the best fit by step 100. The average acceptance fraction for the walkers is 0.51.

\begin{deluxetable}{ccccccc}
\tabletypesize{\footnotesize}
\tablewidth{0pt}
\tablecaption{Data Summary\label{data_table}}
\tablehead{\colhead{Line} & \colhead{Channel Width} & \colhead{Beam Size} & \colhead{Beam PA} & \colhead{rms} & \colhead{Integrated Intensity} & \colhead{Peak Flux} \\ \colhead{} & \colhead{} & \colhead{(FWHM)} & \colhead{} & \colhead{(mJy/beam)} & \colhead{} & \colhead{(Jy/beam)}}
\startdata
CO(2-1) & 0.20 km s$^{-1}$\tablenotemark{a} & 0$\farcs$58x0$\farcs$47\tablenotemark{b} & 75.6$^{\circ}$ & 6.5 & 44.1$\pm$0.3 Jy km s$^{-1}$& 0.53\\
C$^{18}$O(2-1) & 0.21 km s$^{-1}$\tablenotemark{a} & 0$\farcs$58x0$\farcs$49\tablenotemark{b} & -88.8$^{\circ}$ & 1.4 & 5.37$\pm$0.06 Jy km s$^{-1}$& 0.13\\
DCO$^+$(3-2) & 0.21 km s$^{-1}$\tablenotemark{a} & 0$\farcs$59x0$\farcs$50\tablenotemark{b} & 89.6$^{\circ}$ & 1.1 & 1.09$\pm$0.01 Jy km s$^{-1}$& 0.06\\
continuum & 2 GHz & 0$\farcs$42x0$\farcs$35\tablenotemark{c} & -89.1$^{\circ}$ & 0.1 & 626$\pm$19 mJy & 0.13\\ 
\enddata
\tablenotetext{a}{Binned down by a factor of 10 from the original spectral resolution}
\tablenotetext{b}{Derived with robust=0.5}
\tablenotetext{c}{Derived with uniform weighting}
\end{deluxetable}

\begin{turnpage}
\begin{deluxetable}{cccccccccc}
\tabletypesize{\footnotesize}
\tablewidth{0pt}
\tablecaption{DCO$^+$(3-2) model results\label{dco+_results}}
\tablehead{\colhead{Model} & \colhead{$v_{\rm turb}$/c$_s$} & \colhead{Inclination} & \colhead{$R_{\rm in}$} & \colhead{$R_{\rm mid}$} & \colhead{$R_{\rm out}$} & \colhead{log([DCO$^+$/H$_2$]$_{\rm in}$)} & \colhead{log([DCO$^+$/H$_2$]$_{\rm mid}$)} & \colhead{log([DCO$^+$/H$_2$]$_{\rm out}$)} & \colhead{$\chi^2_{\nu}$\tablenotemark{a}} \\ \colhead{} & \colhead{} & \colhead{($^{\circ}$)} & \colhead{(au)} & \colhead{(au)} & \colhead{(au)} & \colhead{} & \colhead{} & \colhead{} & \colhead{}}
\startdata
Fiducial & $<0.04$ & 48.8$^{+0.2}_{-0.1}$ & 65.7$\pm$0.9 & 149.9$^{+0.5}_{-0.7}$ & 259$\pm$1 & -10.83$\pm$0.03 & -10.79$\pm$0.01 & -10.99$\pm$0.01 & 1.1076\\
\hline
Narrow Rings & $<$0.6 & 48.7$\pm$0.2 & 77.9$\pm$0.7 & 151.7$\pm$0.3 & 252.7$^{+0.9}_{-0.7}$ & -9.43$^{+0.05}_{-0.04}$ & -9.64$^{+0.07}_{-0.02}$ & -10.02$^{+0.03}_{-0.01}$ & 1.1084\\
Variable ring width\tablenotemark{b} & $<$0.06 & 48.6$^{+0.1}_{-0.2}$ & 60.6$^{+0.8}_{-1.1}$ & 148.0$^{+1.4}_{-1.3}$ & 266$^{+1}_{-2}$ & -10.86$\pm$0.04 & -11.06$^{+0.04}_{-0.02}$ & -11.15$\pm$0.02 & 1.1074\\
$sys$=1.2 & $<$0.05 & 48.9$^{+0.1}_{-0.02}$ & 66.9$\pm1.1$ & 149.8$\pm0.7$ & 259.7$^{+1.0}_{-0.9}$ & -10.62$\pm$0.04 & -10.62$\pm0.01$ & -10.89$\pm0.01$ & 1.1075\\
$sys=0.8$ & $<$0.04 & 48.7$^{+0.1}_{-0.2}$ & 65.8$^{+1.5}_{-1.3}$ & 149.8$\pm0.7$ & 259$\pm1$ & -11.02$\pm0.03$ & -10.95$\pm0.01$ & -11.11$\pm$0.01 & 1.1076\\
$T_{\rm mid0}$ variable\tablenotemark{c} & $<$0.12 & 48.8$^{+0.1}_{-0.2}$ & 65.9$^{+1.0}_{-0.9}$ & 149.4$\pm0.3$ & 260.6$\pm$1.1 & -10.30$\pm0.07$ & -10.44$^{+0.06}_{-0.04}$ & -10.85$^{+0.03}_{-0.02}$ & 1.1075\\
\enddata
\tablecomments{Median, plus 3$\sigma$ ranges, of the posterior distribution functions derived for each model. No significant degeneracies were found between the listed parameters.}
\tablenotetext{a}{DCO$^+$(3-2) reduced chi-squared for the model defined by the median of the PDFs. Given the large number of degrees of freedom in the data, the differences in reduced chi-squared between the different models are highly significant.}
\tablenotetext{b}{The radial width of the middle and outer ring are no longer fixed at 10\%\ of the radial location, but are instead allowed to vary as free parameters. We derive a FWHM of the middle and outer rings of  60$\pm$3 au and 94$\pm$7 au respectively.}
\tablenotetext{c}{$T_{\rm mid0}$ is allowed to vary, with DCO$^+$(5-4) included in the fitting. $T_{\rm mid0}$ is constrained to 12.2$^{+0.4}_{-0.5}$ K}
\end{deluxetable}
\end{turnpage}
\clearpage

\begin{deluxetable}{cc}
\tabletypesize{\footnotesize}
\tablewidth{0pt}
\tablecaption{C$^{18}$O(2-1) model results\label{c18o_results}}
\tablehead{\colhead{Parameter} & \colhead{Value}}
\startdata
$v_{\rm turb}$ & $<$0.05 c$_s$\\
$\gamma$ & 0.62$\pm$0.01\\
incl & 48.26$^{+0.07}_{-0.05}$\\
$R_{\rm in}$\tablenotemark{a} & 35.6$^{+1.7}_{-2.0}$ au\\
log([CO/H$_2$]) & -4.610$\pm$0.003\\
$R_{\rm break}$ & 102$^{+4}_{-7}$ au\\
$q_{\rm in}$ & -1.4$^{+0.2}_{-0.1}$\\
$\chi^2_{\nu}$\tablenotemark{b} & 0.992\\
\enddata
\tablecomments{Median, plus 3$\sigma$ ranges, of the posterior distribution functions derived from fitting to the C$^{18}$O(2-1) data.}
\tablenotetext{a}{During the final MCMC trial, this parameter was fixed at 35.6 au to facilitate convergence of the other parameters.}
\tablenotetext{b}{Reduced chi-squared for the model defined by the median of the posterior distribution functions}
\end{deluxetable}

\begin{deluxetable}{cc}
\tabletypesize{\footnotesize}
\tablewidth{0pt}
\tablecaption{CO(2-1) model result\label{co_results}}
\tablehead{\colhead{Parameter} & \colhead{Value}}
\startdata
$q$ & -0.27$^{+0.01}_{-0.02}$\\
log($R_c$ (au)) & 2.30$^{+0.03}_{-0.02}$\\
v$_{\rm turb}$ & $<$0.06 c$_s$\\
$T_{\rm atm0}$ & 87$^{+1}_{-2}$ K\\
$T_{\rm mid0}$ & 17.8$^{+0.8}_{-0.6}$ K\\
incl & 47.5$^{+0.5}_{-0.2}$\\
$R_{\rm in}$ & 11$^{+5}_{-3}$ au\\
$R_{\rm break}$ & 70$^{+17}_{-5}$ au\\
$q_{\rm in}$ & -0.57$^{+0.05}_{-0.04}$\\
$\chi^2_{\nu}$\tablenotemark{a} & 1.16\\
\enddata
\tablecomments{Median, plus 3$\sigma$ ranges, of the posterior distribution functions derived from fitting to the CO(2-1) data. }
\tablenotetext{a}{Reduced chi-squared for the model defined by the median of the posterior distribution functions}
\end{deluxetable}

\end{document}